\documentclass[letterpaper,twocolumn,10pt]{article}
\usepackage{usenix2019_v3}
\usepackage{graphicx}

\usepackage{subfigure}
\usepackage{subcaption}
\usepackage{amsmath}
\usepackage{comment}
\usepackage{algorithm}
\usepackage{algorithmic}
\usepackage{url}
\usepackage{diagbox}
\usepackage{upgreek}
\usepackage{dblfloatfix}
\usepackage{multirow}
\usepackage{svg}
\usepackage{amssymb}
\usepackage{hyperref}
\usepackage{breakurl}

\microtypecontext{spacing=nonfrench}
\DeclareUnicodeCharacter{2212}{-}

\begin{document}
\title{\Large \bf PM-Dedup: Secure Deduplication with Partial Migration from Cloud to Edge Servers}

\author{
Zhaokang Ke \\
University of Minnesota \\
\and
Haoyu Gong \\
University of Minnesota \\
\and
David H.C. Du \\
University of Minnesota \\
}

\maketitle

%

%
\begin{abstract}
Currently, an increasing number of users and enterprises are storing their data in the cloud but do not fully trust cloud providers with their data in plaintext form. To address this concern, they encrypt their data before uploading it to the cloud. However, encryption with different keys means that even identical data will become different ciphertexts, making deduplication less effective. Encrypted deduplication avoids this issue by ensuring that identical data chunks generate the same ciphertext with content-based keys, enabling the cloud to efficiently identify and remove duplicates even in encrypted form. Current encrypted data deduplication work can be classified into two types: target-based and source-based. Target-based encrypted deduplication requires clients to upload all encrypted chunks (the basic unit of deduplication) to the cloud with high network bandwidth overhead. Source-based deduplication involves clients uploading fingerprints (hashes) of encrypted chunks for duplicate checking and only uploading unique encrypted chunks, which reduces network transfer but introduces high latency and potential side-channel attacks,  which need to be mitigated by Proof of Ownership (PoW), and high computing overhead of the cloud. So, reducing the latency and the overheads of network and cloud while ensuring security has become a significant challenge for secure data deduplication in cloud storage. In response to this challenge, we present PM-Dedup, a novel secure source-based deduplication approach that relocates a portion of the deduplication checking process and PoW tasks from the cloud to the trusted execution environments (TEEs) in the client-side edge servers. We also propose various designs to enhance the security and efficiency of data deduplication. 
\end{abstract}

\section{Introductions}\label{sec:introduction}

Cloud storage services (such as Dropbox, Google Drive, OneDrive, iCloud Drive, etc.) have become very popular in recent years for storing and sharing data. The Google 2023 Data and AI Trends Report states that by 2026, the global data generation rate will reach 7 petabytes per second, with only 10$\%$ being original data and the remaining 90$\%$ being replicated data. To handle the vast amounts of data, many cloud storage services utilize deduplication technology to eliminate duplicate content and optimize storage space. The common approach of data deduplication is to eliminate redundant data and store only one physical copy called unique data. This physical copy can be referenced and accessed by a file/object using small-size references. According to recent publications, implementing cross-user deduplication can result in substantial space savings, with more than 50$\%$ in primary storage  ~\cite{shamma2011capo} and up to 90$\%$ to 95$\%$ in backup storage ~\cite{wallace2012characteristics} ~\cite{shin2017survey}.

Facing the increasing data privacy concerns and security threats, clients are becoming more cautious about the safety of their data and are increasingly reluctant to trust cloud storage service providers with their unencrypted data. This concern leads them to encrypt their data before uploading it to the cloud. However, encrypting the same data with different keys results in distinct ciphertexts, which reduces the effectiveness of deduplication after encryption ~\cite{shukla2023cisco}. To address this issue, encrypted deduplication techniques have been developed which use a hash of the data content itself as a key for encryption ~\cite{hasan2005evolution, kotla2007safestore, vrable2009cumulus, wallace2012characteristics}. In this method, clients encrypt the original plaintext data using content-derived keys to produce ciphertexts. This approach allows identical data to be encrypted into the same ciphertext when using the same key, thereby enhancing the effectiveness of encrypted deduplication by enabling the cloud to perform duplication checks based on these identical ciphertexts.

Existing encrypted deduplication can be categorized into two approaches: 1) Target-based encrypted deduplication, where clients upload all ciphertext chunks, even if they are duplicates, to the cloud for deduplication. This approach results in additional network bandwidth for uploading duplicate ciphertext chunks, leading to increased costs related to data transfer. 2) Source-based encrypted deduplication, where clients compute the fingerprints of the ciphertext chunks and upload them to the cloud for a duplicate check, and then only upload unique ciphertext chunks. The cloud performs the duplication checks by comparing the fingerprints of the ciphertext chunks. Since the same plaintext encrypted with the same key will result in identical ciphertexts, their fingerprints will also be identical, allowing the cloud to identify duplicates or unique data. While this method can reduce data transfer, it introduces three potential problems.

Firstly, it is vulnerable to side-channel attacks. One possible attack occurs when a compromised client generates and sends guessed fingerprints of ciphertext chunks to the cloud. If the cloud confirms the existence of these chunks and does not require an upload, the client can infer that the data is already stored in the cloud, potentially breaching data confidentiality. To counter this, clients must prove to the cloud that they own the data chunks through a Proof of Ownership (PoW) protocol~\cite{halevi2011proofs, harnik2010side, di2012boosting, di2016proof}. PoW can be executed through a challenge-response protocol. That is, when a client does not need to upload a data chunk, the cloud generates a challenge and a response based on the chunk already stored on the cloud and sends the challenge to the client to demand a response. The client must use its original chunk to produce a response and send it back to the cloud. The cloud then can verify the ownership by matching the client's response with its pre-computed response. However, this protocol increases latency due to the additional round-trip challenge-response communication and the time required for the cloud to generate the challenge and response in real-time.

Secondly, it introduces high latency due to the additional deduplication checking step required at the cloud before uploading data. This means clients must wait for the cloud to perform a duplicate check on the uploaded fingerprints, resulting in a delay before they can upload the actual data chunks or face the PoW challenge.

Lastly, this approach results in significant overhead in the cloud, including the costs associated with duplicate checks, ownership verification, and the real-time computation of challenges and responses for PoW. These operations require considerable resources, and as the number of client queries increases, the associated overhead can become a substantial bottleneck.

The advancements in trusted execution environments (TEEs) ~\cite{sabt2015trusted, jauernig2020trusted, kang2021iceclave}, which provide a secure execution environment for processing sensitive code and data with confidentiality, availability, and integrity guarantees, offer new possibilities for enhancing the performance of encrypted deduplication. Currently, several TEEs across different platforms are commonly used, including ARM TrustZone, AMD SEV, Intel SGX, Sanctum, and Sanctuary, among others~\cite{jauernig2020trusted}. In this paper, we focus on one specific TEE: Intel SGX ~\cite{kim2019shieldstore, harnik2018securing, wang2023svtpm, oh2020trustore}, which is popular and easy to deploy. It allows for the allocation of an isolated memory region (enclave) against the host system, attests in-enclave contents via remote attestation and can securely move in-enclave contents into unprotected memory via encryption ~\cite{ren2021accelerating, yang2022secure}. Given the considerable security benefits of SGX, as confirmed by recent research, we are motivated to migrate a portion of the deduplication check process and the PoW tasks from the cloud to the enclave of an edge server in client side. This approach aims to reduce the computation cost and the latency due to the extra checking step and proof-of-ownership, while still ensuring security, bandwidth efficiency, and storage optimization.

We introduce PM-Dedup, designed specifically for a single organization or company with multiple branches or offices spread across various geographical regions. In this scenario, the cloud system maintains a list of fingerprints uploaded by the organization's clients. These fingerprints are derived from the encrypted data chunks of the files that clients upload to the cloud. Each geographical location within the organization is equipped with its own edge server. This structure inherently increases the likelihood of duplicate data chunks being uploaded, as clients within the same organization often share common data. PM-Dedup proposes migrating tasks originally performed on the cloud, such as deduplication checks and PoW, to the edge server closer to the client, where the visit latency is significantly lower compared to that of the cloud. To achieve this, the cloud pre-computes challenges and responses for selected chunks and files, rather than generating them in real-time. These pre-computed challenges and responses are then transferred to the edge server, enabling faster PoW verification. Additionally, by selecting and transferring a subset of fingerprints (the share-index) to the edge server for deduplication checks, PM-Dedup further reduces the need for cloud interactions, thereby decreasing latency and lowering the cloud’s overhead.

While realizing this approach, we encounter two challenges. First, it is crucial to ensure the security of the information shared by the cloud with the edge servers, including the share-index for deduplication checks and pre-computed challenges and responses for PoW.
This is because cloud systems are typically reluctant to share their data in an insecure manner, making the safeguarding of these elements from potential attacks paramount. Second, the selection of an efficient share-index, as well as pre-computed challenges and responses are very important. Because careful selection of these elements increases the likelihood of handling these tasks on the edge server, thereby reducing communication latency, network traffic, and overhead on the cloud. To this end, we propose four major components of PM-Dedup:

\begin{itemize}
    \item \textbf{Efficient Share-Index Generation:} Optimizes deduplication by accurately identifying both current and future high-impact fingerprints from the cloud.

    \item \textbf{Dual-Level Lightweight and Responsive PoW:} Ensures efficient ownership verification by combining file- and chunk-level PoW. Pre-computed challenges and responses are utilized to reduce the response time for PoW verification.

    \item \textbf{SGX-based Migration:} Securely migrates critical operations from cloud to Intel SGX enclaves on the edge server, minimizing overhead on the cloud and latency.

    \item \textbf{Optimized Edge Server Management:} Dynamically manages the share-index, along with pre-computed challenges and local fingerprints, to reduce computational overhead.
\end{itemize}

The rest of the paper is structured as follows: Section \ref{sec:background} provides the necessary background and motivation. Section \ref{sec:overview} presents an overview of our design. Section \ref{sec:design issue} describes the detailed design. Sections \ref{sec:implementation} and \ref{sec:evaluation} discuss the implementation details and evaluation results respectively. Finally, Section \ref{sec:conclusion} concludes the paper.

\section{Background $\&$ Motivation}\label{sec:background}

\subsection{Data Deduplication} \label{sec:dedup}
We focus on chunk-based deduplication of each file/object.~\cite{meyer2012study, wallace2012characteristics, zhu2008avoiding}, a pivotal technique in data storage management that optimizes storage efficiency by eliminating redundant data. The process involves partitioning data into chunks and assigning each chunk a unique cryptographic hash-based fingerprint based on the content of the chunk, ensuring that different chunks will have distinct fingerprints. Because different chunks generate different fingerprints, even slight variations in the data result in unique identifiers. Consequently, these fingerprints can facilitate the detection and removal of duplicates.

In the deduplication system, each unique chunk is stored only once, while duplicate occurrences are replaced with a reference to the original. To manage this process, the system maintains a fingerprint index, a key-value store mapping fingerprints to the physical addresses of their corresponding chunks. Additionally, the system stores a file recipe for each file, listing references to all its constituent chunks. This recipe aids in future file reconstruction. 

\subsection{Encrypted Deduplication} \label{sec:encrypted deduplication}
As more users and enterprises increasingly prioritize data privacy and security, the necessity of encrypting data before storage becomes paramount~\cite{yang2020data, wang2010secure, kamara2010cryptographic}. While conventional data deduplication offers significant storage benefits, it encounters difficulties when applied to encrypted data. The reason for this lies in the nature of encryption itself: same data encrypted with different keys yields different ciphertexts, which makes traditional deduplication techniques ineffective due to they have different fingerprints. 

\textbf{Message Locked Encryption (MLE)}: To overcome this obstacle, Message Locked Encryption (MLE) was introduced ~\cite{bellare2013message}. MLE is an encryption scheme that generates the encryption key based on the content of the message itself. Therefore, identical data chunks will always produce the same key to be used for encryption, resulting in identical ciphertexts. This approach allows for effective deduplication of encrypted data since identical plaintexts—even when independently encrypted—will yield identical ciphertexts. Hence, MLE enables deduplication systems to recognize and eliminate duplicate encrypted data.

\textbf{Brute-force attacks in MLE:} However, MLE is susceptible to brute-force attacks because the encryption key is directly derived from the plaintext chunk. Specifically, an adversary can infer the input plaintext chunk from a target ciphertext chunk by systematically generating the MLE keys for all possible plaintext chunks and checking if any generated plaintext chunk produces the target ciphertext when encrypted ~\cite{ren2021accelerating}.

Server-aided MLE enhances the security of encrypted deduplication against offline brute-force attacks by utilizing a dedicated key server for MLE key generation~\cite{keelveedhi2013dupless}. When encrypting a plaintext chunk, the client sends the fingerprint of the plaintext chunk to the key server, which uses the fingerprint and a global secret to generate the MLE key. This key is then securely returned to the client who will use it to encrypt the plaintext chunk. The security of this method hinges on the global secret's confidentiality. If compromised, the security reverts to traditional MLE's level. To further safeguard the process, server-aided MLE employs an oblivious pseudo-random function (OPRF)~\cite{naor2004number} to blind the fingerprints, ensuring the key server cannot learn them. Additionally, it rate-limits key generation requests to mitigate online brute-force attacks.

 \textbf{Side-channel attack}: In contrast to target-based encrypted deduplication which involves clients uploading all ciphertext chunks to the cloud and, therefore, suffers from high bandwidth requirements~\cite{yang2022secure}, our research focuses on source-based encrypted deduplication. In this method, clients compute the fingerprints of the ciphertext chunks and upload them to the cloud for a duplicate check before uploading the actual ciphertext chunks. Only unique ciphertext chunks are then uploaded to the cloud which is beneficial in reducing data transfer and associated costs. However, it is vulnerable to side-channel attacks that pose substantial security risks ~\cite{halevi2011proofs, harnik2010side, di2012boosting, di2016proof}. Two main forms of side-channel attacks exist, both of which can be carried out by compromised clients. The first form of attack allows the compromised client to verify the existence of a particular ciphertext chunk (potentially corresponding to a sensitive piece of information like a password) by dispatching the fingerprint of the anticipated ciphertext chunk to the cloud. This strategy enables the compromised client to extract sensitive information belonging to other clients, thereby breaching confidentiality. The second form of attack involves the compromised client gaining unauthorized access to stored chunks of other clients. By using the fingerprint of a target ciphertext chunk, the attacker can convince the cloud that it is the legitimate owner of the corresponding ciphertext chunk, thus gaining full access rights.

\textbf{Proof of Ownership (PoW)}: To remedy the security threats mentioned above, the PoW protocol is proposed to ensure that a client can demonstrate its ownership of data before gaining access to duplicate data already stored in the cloud. 

The fundamental goal of PoW is to prevent unauthorized
access to data by verifying that a client requesting access to data is indeed its rightful owner. PoW can be executed through a challenge-response protocol. When a client uploads a fingerprint of a duplicated ciphertext, the cloud generates a challenge and a response based on the data already stored on the cloud and sends the challenge to the client. The client must then generate a response based on the challenge and the data itself and send it back to the cloud. The cloud verifies ownership by matching the client's response with its pre-computed response. If the responses match, the client is confirmed as the rightful owner of the data. This process ensures that the client possesses the actual data and prevents a compromised client from accessing data it does not own.

Data can be verified at two granularity levels: file-level and chunk-level. File-level validation involves verifying the entire file's ownership which, if successful, confirms ownership of all chunks within the file. This approach is quicker and less resource-intensive but can be too coarse. For instance, if even a small part of the file changes, the cloud may not find the corresponding pre-computed challenge, causing the entire file to fail the PoW and requiring the upload of the whole file despite many chunks being duplicates that already exist in the cloud. These duplicates, which could have been handled through source-based deduplication to save network costs, must be uploaded due to the PoW failure. This scenario contradicts the intent of source-based deduplication. On the other hand, chunk-level validation, which is finer-grained, is preferred in most source-based deduplication systems although this approach incurs higher overhead compared to file-level validation. 

\textbf{Drawbacks of Source-Based Deduplication}: 
While source-based deduplication reduces data transfer, it introduces two major limitations compared to target-based deduplication. First, PoW and deduplication checks require frequent round-trip interactions with the cloud, resulting in higher latency. Additionally, the real-time generation of challenges and responses for PoW further exacerbates these delays. Second, it increases the cloud's overhead, as the cloud must continuously manage PoW verification, deduplication checks, and the computation of challenges and responses.

\subsection{Trusted Execution Environments}\label{sec:TEE}
A Trusted Execution Environment (TEE) is a secure processing environment that operates on a separation kernel, providing a robust defensive front against both software and physical attacks. The TEE upholds the authenticity of the executed code, maintains the integrity of the runtime states—including CPU registers, memory, and sensitive I/O—and ensures the confidentiality of its code, data, and runtime states when stored in persistent memory. Furthermore, it is equipped with the capability to provide remote attestation, thereby affirming its trustworthiness to third parties \cite{sabt2015trusted}. Currently, several TEEs across different platforms are commonly used, including ARM TrustZone, AMD SEV, Intel SGX, Sanctum, and Sanctuary, among others~\cite{jauernig2020trusted}. Our research primarily centers on SGX, a prominent example of TEE. Intel SGX incorporates three key features:

\begin{itemize}
    \item \textbf{Isolation:} SGX offers the ability to designate an isolated memory region—referred to as an enclave—that is safeguarded against the host system. This functionality ensures that the execution of code and data within the enclave remains protected against any potential threats originating from the host.

    \item \textbf{Attestation:} SGX is capable of verifying the authenticity and integrity of contents within an enclave using a process known as remote attestation. This process allows a remote entity (e.g., a cloud service) to validate that the enclave's contents are trustworthy and have not been tampered with.

    \item \textbf{Sealing:} Intel SGX provides the capacity for an enclave to securely transfer its contents into the unprotected memory. This process, known as sealing, involves the encryption of data before its eviction from the enclave, providing an additional layer of protection.
\end{itemize}

\textbf{Limitations of Intel SGX}: While the SGX provides robust security features, it has two primary drawbacks:
\begin{itemize}
    \item \textbf{Memory Constraint:} The SGX enclave page cache (EPC) provides a restricted space for protected code and data. If the enclave's data exceeds the EPC, it necessitates encryption and eviction of surplus pages to unprotected memory.

    \item \textbf{Performance Overhead:} Interactions between applications and the enclave introduce a context-switching overhead. This overhead arises when applications issue enclave calls (ECalls) to securely access in-enclave contents.
\end{itemize}

\section{PM-Dedup Overview} \label{sec:overview}

\subsection{Design Principles} \label{sec:principle}
PM-Dedup aims to address several challenges associated with source-based deduplication, and its design is guided by the following principles:

 \begin{itemize}
    \item \textbf{Confidentiality:} PM-Dedup aims to ensure the security of both client-uploaded data and cloud-shared data. Accordingly, clients need to encrypt their data using server-aided MLE, which helps prevent data leakage and resist brute-force attacks. Additionally, PoW mechanisms are required to protect against side-channel attacks from clients. Furthermore, data shared by the cloud, such as share indexes and pre-computed challenges, must be securely stored and managed to prevent unauthorized access.

    \item \textbf{Response Efficiency:} PM-Dedup aims to optimize system responsiveness by performing most of the deduplication checks and PoW tasks at the edge server side, where latency is much lower than at the cloud. Additionally, it seeks to enhance the speed of PoW real-time responses, further reducing response time.

    \item \textbf{Bandwidth Efficiency:} PM-Dedup is intended to identify and eliminate redundant data before it is sent to the cloud, prioritizing bandwidth conservation in the same manner as source-based deduplication.

    \item \textbf{Computational Efficiency:} PM-Dedup seeks to employ lightweight share-index selection and PoW mechanisms to reduce overall system overhead, thereby improving the efficiency and scalability of the entire system.
\end{itemize}






\subsection{Overview Architecture} \label{sec:over arc}

\begin{figure}[!t] 
\centering 
\includegraphics[width=0.5\textwidth]{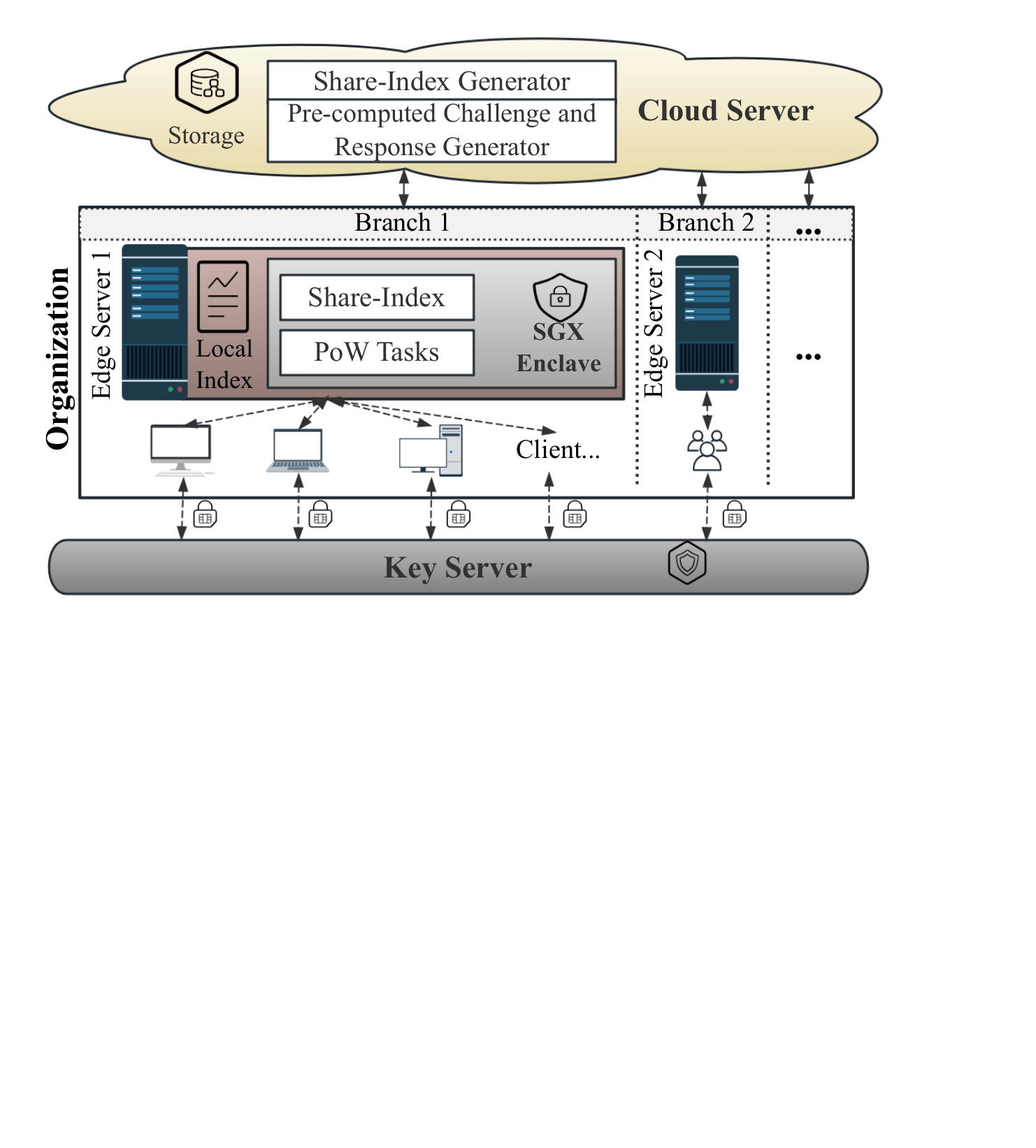} 
\caption{PM-Dedup architecture.} 
\label{PM-Dedup architecture} 
\end{figure}

As shown in Figure~\ref{PM-Dedup architecture}, the PM-Dedup system is designed for a single organization with branches distributed across different geographical regions. Each branch is supported by a dedicated edge server equipped with an SGX enclave. The system also includes a centralized cloud server and a trusted third-party key server, which are shared across all branches. Since clients within the same organization often share similar data, they are likely to upload duplicate data across different branches.

\textbf{Key Server}: PM-Dedup adopts a server-aided MLE scheme. The key server, a trusted third-party entity, is centrally located to serve all branches of the organization. In this scheme, the client sends a fingerprint (hash) of the plaintext chunk to the key server. The key server, which holds a global secret unknown to the clients, uses this fingerprint along with the global secret to generate a secure MLE key. The generated MLE key is then securely returned to the client, who uses it to encrypt the plaintext chunk. Note that, if a key server becomes a performance bottleneck, it can be duplicated in multiple locations as shown in some recent works which  employ distributed key servers to enhance scalability and reduce latency~\cite{duan2014distributed, miao2015secure}, our focus is not on optimizing key server architecture. Consequently, to simplify our discussion, we utilize a single centralized key server, as shown in Figure~\ref{PM-Dedup architecture}.

\textbf{Cloud Server}: The cloud server primarily stores the encrypted data and handles basic client requests such as read, update, and delete operations. The cloud server also manages a full-index, which is a comprehensive collection of all fingerprints from different clients across different branches of the organization, encompassing all the data required for deduplication checks.

To enhance deduplication efficiency, the cloud server employs a share-index generator as shown in Figure~\ref{PM-Dedup architecture}. This generator selects a subset of high-frequency and logically related fingerprints from the full-index to create the Share-Index. The Share-Index is then shared with the edge servers and securely stored in the SGX enclave of the edge servers. The purpose of the Share-Index is to provide a global perspective, ensuring high hit ratios during the duplicate check process at each edge server.

In addition, the cloud server incorporates a pre-computed challenge and response generator, also depicted in Figure~\ref{PM-Dedup architecture}. Unlike real-time generation, this pre-computed challenge and response generator is responsible for pre-generating challenges and responses in advance for selected files and chunks that are most likely to be accessed frequently. These pre-computed challenges and responses are then migrated to the edge servers, where they are securely stored in the SGX enclave of each edge server. These pre-computed challenges and responses are used by the PoW protocol.

By transmitting the pre-computed challenges, responses, and the Share-Index to the SGX enclaves of the edge servers, PM-Dedup ensures the security of the shared information while ensuring that the most relevant and frequently accessed data is readily available for deduplication  and PoW checks at the local edge side. Thus, the clients can minimize the need for frequent communication with the cloud server, where latency is much higher compared to the communication delays with edge servers, thereby decreasing overall latency and reducing cloud overhead.

\textbf{Edge Server}: As shown in Figure~\ref{PM-Dedup architecture}, each edge server, dedicated to a specific geographical location or branch of the organization, leverages an SGX enclave to securely manage the pre-computed challenges, responses, and the Share-Index received from the cloud server. Given the constraints of the limited SGX enclave memory size and the high cost of SGX enclave interactions, the edge server also maintains a fixed-size local-index outside the enclave. This local-index stores both file and chunk fingerprints, managed using an LRU (Least Recently Used) strategy, to reflect the most recently used files and chunks from the clients in the branch it serves.

When a client intends to store data in the cloud, the process begins with the client chunking files into plaintext data chunks. To ensure data privacy, the client interacts with the Key Server to obtain encryption keys by sending fingerprints of these data chunks. The client then encrypts each chunk using these keys, preserving data integrity from the onset.

Before performing source-based deduplication process, a client must prove its ownership of the data. This is facilitated by the challenge-response based PoW method detailed in Section \ref{sec:pow}. Here, we use chunk-level PoW to illustrate the process. The SGX enclave in the edge server manages a set of selected pre-computed challenges and responses from the cloud for PoW at the edge server. The client first uploads the fingerprint of the ciphertext chunk to the edge server. The edge server then identifies an unverified match of the fingerprint of the ciphertext chunk. It then send the pre-computed challenge of the ciphertext to the client. The client must respond to this challenge and send the corresponding response to the edge server for verification. If the response is correct, the edge server verifies the client's ownership of the chunk. If the edge server does not have the required pre-computed challenge (i.e., do not have a match), the validation request is escalated to the cloud server for verification. Upon successful validation, which confirms that the client possesses the original data chunk, the edge server proceeds with the source-based deduplication process.

 Once the ownership has been verified, the chunk's fingerprints are uploaded to the edge server for the duplicate check. Here, a systematic tiered search is initiated: the local-index is the primary checkpoint. If no match is found, the system escalates to the Share-Index stored in the SGX enclave. If the chunk remains unmatched even at this level, the final verification occurs with the cloud server. For chunks deemed unique after all checks, the client will upload them to the cloud. Once uploaded, the necessary updates are made: the local-index, the Share-Index, and relevant cloud records are updated to reflect this new chunk, ensuring the system remains primed for subsequent deduplication checks.

The reason we need the share-index instead of relying solely on the local-index:

\textbf{(i) Increased Deduplication Efficiency and Reduced Cloud Interaction}

\textit{Local-index Limitations:} The local-index on the edge server is limited to the historical uploads to that particular edge server. This narrow scope may miss duplicates that are common across different branches in the same organization, leading to lower deduplication hit ratios.

\textit{Broader Scope:} The Share-Index, generated by the cloud server, includes fingerprints of high-frequency and logically related chunks from a global perspective. This ensures a higher likelihood of detecting duplicate at the edge server, thereby reducing the need for high-latency interactions with the cloud server.

\textbf{(ii) Adaptability to Data Trends}

\textit{Dynamic Data Patterns:} Data access patterns and redundancy levels can vary over time. The local-index might not adapt quickly to these changes, leading to inefficiencies.

\textit{Periodic Updates:} The Share-Index is periodically updated based on deduplication trends and access patterns observed at the cloud server. This dynamic updating ensures that the edge servers have the most relevant and high-impact fingerprints, maintaining high deduplication efficiency even as data patterns evolve.


\subsection{Threat Model} \label{sec:threat model}
In this section, we detail the threat model and security assumptions of PM-Dedup.

 \paragraph{\textbf{Honest-but-Curious Cloud Server:}}
 
The cloud server is trusted to perform its operations correctly, such as managing storage and handling deduplication processes \cite{chai2012verifiable}. However, it is considered curious, meaning it may attempt to infer information from the data it stores. To protect against potential data leakage, encryption is performed on the client side, ensuring that the cloud never handles plaintext data.

\paragraph{\textbf{Partially Trusted Edge Server:}}

The edge server is trusted to manage tasks such as handling the share-index, performing deduplication checks, and executing PoW protocol \cite{malekzadeh2021honest, sun2022privacy}. However, it is vulnerable to data leakage due to potential physical attacks, software exploits, or curiosity about the data shared by the cloud. To mitigate these risks, critical operations and sensitive data from the cloud are securely confined within an SGX enclave on the edge server.

\paragraph{\textbf{Untrusted Client:}}

The client is not trusted and could be compromised or act maliciously. The system assumes that a client might seek unauthorized access to original data chunks residing on the cloud or the edge server. To counter these threats, PoW protocols are used to verify the client's ownership of data before allowing deduplication or access to stored data.

PM-Dedup operates under the following security assumptions:

\paragraph{\textbf{Secure SGX Operation:}} 
The SGX enclave operates as intended, providing effective isolation, attestation, and sealing mechanisms. Despite the known limitations of SGX such as vulnerability to denial-of-service or side-channel attacks, the crux of our research centers on migrating the secure deduplication process from the cloud to the edge server with the aim of reducing latency and query costs. We acknowledge that other TEEs could be utilized to mitigate these attacks. However, our focus is placed on SGX due to its ease of deployment and the simplicity it offers when comparing our work with other related research in the field.

\paragraph{\textbf{Resilient MLE Key:}} We assume the resilience of the MLE key to brute-force attacks. While we acknowledge the potential for such attacks, we trust in the robustness of the key generation process and rate-limiting measures to protect against them.


\section{Design ISSUES} \label{sec:design issue}
PM-Dedup's performance is closely tied to how much of the workload can be efficiently handled by the edge server. This efficiency is primarily  influenced by the selection of the share-index and the effectiveness of the PoW mechanism. In this section, we break down the design issues: Section \ref{sec:share-index selection} introduces the strategies for selecting the share-index in different scenarios; Section \ref{sec:pow} details the design of a lightweight and responsive PoW; and Section \ref{sec:edge} discusses the management of the local index and secure enclave operations at the edge server.

\subsection{Share-index Selection} \label{sec:share-index selection}

As mentioned before, we have a full-index maintained at the cloud level, containing all fingerprints of stored data chunks from different clients across various branches in the organization. Additionally, each edge server, which serves a specific branch within an organization at different geographical locations, maintains a local-index. This local-index reflects the historical uploads from all clients within that specific branch. While crucial for deduplication checks within a single branch, the local-index is limited by its scope, which may miss duplicates common across different branches within the organization.

To address these limitations, we introduce the share-index, which is generated by the cloud server and serves all edge servers within the organization. The share-index is a subset of the full-index, including fingerprints of high-frequency and logically related chunks, selected to further accelerate the deduplication process at the edge server level. As previously mentioned, branches within the same organization share similar interests. Consequently, the same Share-Index is distributed to edge servers across these branches. While, in reality, branches may have different interests, and we could group branches with shared interests and provide them with a common Share-Index, but for simplicity, we assume a uniform Share-Index across all branches in this model.

We face two scenarios: one requires frequent share-index updates on edge servers, necessitating a lightweight selection method. The other involves infrequent updates, challenging us to ensure sustained share-index effectiveness. In this section, we first present a lightweight share-index selection approach based on frequency, using Count-Min Sketch (CMS). Subsequently, we propose a second approach that combines this method with logical locality analysis to enhance long-term performance.

\textbf{\textit{Count-Min Sketch Based Light-weight Selection:}}

\begin{algorithm}[!t]
\caption{CMS-Based Light-weight Selection}
\label{alg:cms}
\begin{algorithmic}[1]

\STATE /* Initialize Count-Min Sketch */
\STATE \textbf{function} \textsc{CMS\_Init}($d$, $w$)
\STATE \ \ \ \ Initialize matrix CMS[$d$][$w$] to zeros
\STATE \ \ \ \ Choose $d$ hash functions $h_1, h_2, \ldots, h_d$
\STATE \textbf{end function}

\STATE /* Add chunk to CMS */
\STATE \textbf{function} \textsc{CMS\_Add}($chunk$)

\STATE \ \ \ \ \textbf{for} $i = 1$ \textbf{to} $d$ \textbf{do}
\STATE \ \ \ \ \ \ \ \ $CMS[i][h_i(chunk)] \leftarrow CMS[i][h_i(chunk)] + 1$
\STATE \ \ \ \ \textbf{end for}

\STATE \textbf{end function}

\STATE /* Estimate chunk frequency */
\STATE \textbf{function} \textsc{CMS\_Frequency}($chunk$)
\STATE \ \ \ \ \textbf{return} $\min\{CMS[i][h_i(chunk)] \text{ for } i = 1 \text{ to } d\}$
\STATE \textbf{end function}

\end{algorithmic}
\end{algorithm}

Frequent updates to the share-index on the cloud server can introduce significant overhead, particularly when tracking high-frequency chunks. This necessitates the need for a lightweight selection method. Traditional methods such as exact counting, hash tables, and heap-based methods are not ideal due to their high memory and computational requirements. Exact counting requires storing a counter for each unique item, leading to significant memory overhead. Hash tables, while efficient for lookups, can become large and unwieldy with high data volumes. Heap-based methods, though useful for maintaining top-k elements, are also memory-intensive and computationally expensive.

We choose Count-Min Sketch (CMS)~\cite{cormode2005improved} which offers a space-efficient probabilistic alternative that addresses these concerns. It approximates the frequency of data chunks using a matrix of counters, making it suitable for environments where memory efficiency and quick updates are crucial. CMS provides a compact representation of frequencies with a controlled error rate, allowing us to efficiently track high-frequency chunks.

Share-index selection process involves aggregating fingerprints from clients across various geographical branches. As shown in Algorithm \ref{alg:cms}, the CMS is implemented by allocating a matrix of counters with \( d \) rows, each row corresponding to a hash function, and \( w \) columns representing the sketch width. The values of \( d \) and \( w \) are chosen based on the desired balance between accuracy and memory usage. Each incoming chunk's fingerprint is hashed \( d \) times, with the corresponding counters incremented across the matrix. 

When tracking the frequency of a chunk, we compute the hash values again and take the minimum value out of the $d$ counters as the estimated frequency. This process enables us to identify high-frequency chunks efficiently. We optimize CMS's accuracy by choosing hash functions that minimize collisions and by setting $w$ and $d$ based on the expected number of unique chunks. This method, while not exact, provides a rapid means of identifying high-frequency chunks, and the memory overhead is significantly reduced. It allows us to adapt to real-time data patterns and maintain performance across the organization's different branches.

\textbf{\textit{Logical Locality Based Chunk Selection:}}

The CMS provides a quick and space-efficient means of identifying high-frequency chunks. However, it focuses solely on frequency and may overlook chunks that are logically related but not currently exhibiting high access frequencies. Logical locality refers to the likelihood of chunks being accessed together due to their contiguous arrangement within a file, based on the spatial and sequential correlation of data chunks in storage systems. While these logically related chunks may not be frequently accessed now, they are likely to be accessed together with high-frequency chunks in the future due to logical access patterns. By considering logical locality, we can enhance deduplication efficiency by capturing relationships between data chunks that may not be immediately apparent from frequency analysis alone.

Logical locality is determined using file recipes, which detail the sequence of chunks within files and are stored on the cloud along with the files. The file recipes we analyze come from files already uploaded by clients across different branches within the same organization. By analyzing these cloud-stored file recipes, we can identify chunks that are often accessed together and prioritize them in the share-index.

As shown in Algorithm \ref{alg:logical}, to quantify logical locality, we introduce a proximity scoring mechanism based on file recipes. This score measures the nearness of each chunk to those identified as high-frequency by the CMS on the cloud side and by the LRU in the local index on the edge server side. The proximity is determined by the chunk's index within its file recipe, recognizing that chunks closely indexed are more likely to be accessed together. Consequently, the distance — the number of intervening chunks — inversely informs the proximity score, ensuring that closer chunks receive a higher score and thus a higher priority for share-index inclusion. Moreover, our algorithm's aggregate scoring mechanism considers a chunk's occurrence across diverse file recipes. The resultant cumulative score is indicative of a chunk's overarching relevance within the dataset of the organization.

After completing these steps, we combine the fingerprints of the chunks identified through CMS with those prioritized by logical locality analysis into the share-index.


\begin{algorithm}[!t]
\caption{Logical Locality Based Chunk Selection}
\label{alg:logical}
\begin{algorithmic}[1]
\STATE \textbf{Input:} Frequent Chunks $F$, File Recipe Collection $R$, Proximity Threshold $T$
\STATE \textbf{Output:} Share-Index Selection $S$

\STATE $ScoreMap \leftarrow$ empty map to hold cumulative proximity scores for candidates
\FORALL{frequent chunk $f \in F$}
    \FORALL{file recipe $r \in R$ that contains $f$}
        \STATE $i_f \leftarrow$ index of $f$ in file recipe $r$
        \FORALL{chunk $c$ in file recipe $r$}
            \STATE /* Calculate the distance of $c$ from $f$ in $r$ */
            \STATE $d_{fc} \leftarrow |$index of $c$ in $r - i_f|$
            \STATE /* Update cumulative score */
            \STATE $ScoreMap[c] \leftarrow ScoreMap.get(c, 0) + \frac{1}{1 + d_{fc}}$
        \ENDFOR
    \ENDFOR
\ENDFOR

\STATE $S \leftarrow$ Top elements from $ScoreMap$ based on score
\RETURN $S$
\end{algorithmic}
\end{algorithm}

\subsection{Dual-Level Lightweight and Responsive PoW:}\label{sec:pow}
By selecting the share-index from the cloud and securely storing it on the edge servers, we can offload certain deduplication check tasks from the cloud to the edge server. This strategy prompts
us to investigate whether we can similarly offload a portion of PoW tasks
from the cloud to edge servers, where latency is lower. However, PoW requires the computation of challenges and responses, which typically demands significant computational resources. Since edge servers have limited computing power, they typically cannot perform these calculations in real time ~\cite{wang2019edge, luo2021resource}. We address this challenge by having the cloud server pre-computes challenges and responses for chunks and files that are likely to be frequently accessed on the edge server. These pre-computed challenges and responses are then stored in the SGX enclave of the edge server, allowing it to focus solely on verification tasks. This approach reduces the edge server's computational burden, making PoW offloading feasible, while also decreasing response time by eliminating real-time computation. The pre-computation can be performed during periods of lower resource demand, further optimizing the system's efficiency.

Regarding the granularity of data verification, as discussed in Section ~\ref{sec:encrypted deduplication}, there are two levels: file-level and chunk-level. The file-level PoW verification is quick and resource-efficient but operates on coarse granularity, while chunk-level PoW verification offers finer granularity but incurs higher overhead. This trade-off motivated us to first employ file-level PoW to efficiently eliminate a broad set of duplicates. After this initial filtering, we applied chunk-level PoW to the remaining data, enabling us to capture finer-grained duplicates.

In this section, we introduce a lightweight and responsive PoW scheme that assigns pre-computed challenges and responses from the cloud server to the edge server, enabling ownership verification at both file and chunk levels. This approach is based on the assumption that it's highly unlikely for a malicious client to correctly produce \( \textit{K} \) bits of a file or a chunk, especially when each bit is selected from random positions\cite{di2012boosting}. Moreover, the value of \( \textit{K} \) for ownership verification is directly proportional to the file or chunk size.

\textbf{\textit{Data Structure:}}

In the PM-Dedup system, both the edge and cloud servers utilize a dual-structured approach to manage file-level and chunk-level challenge and response. This is achieved through two specialized hash-maps: \textit{F} and \textit{C}. The first hash-map \textit{F} is keyed by the file hash, meaning that each file hash corresponds to a unique entry in the hash-map. This entry includes a file pointer (\textit{ptr}) that directs to the file's location in the storage system, an array of pre-computed challenge/response pairs (\textit{fileChal/fileRes[]}), and an index counter (\textit{idc}) that tracks the number of generated challenges. The \textit{idc} ensures that each challenge remains unique, maintaining the integrity of the PoW process. When a file hash is used as a key, the \textit{F} hash-map quickly retrieves the associated entry, facilitating efficient file-level PoW verification.

Similarly, the \textit{C} hash-map is keyed by the chunk hash. Each chunk hash corresponds to a unique entry in the hash-map, which includes a chunk pointer (\textit{cptr}) directing to the chunk's location, an array of pre-computed chunk-level challenge/response pairs (\textit{chunkChal/chunkRes[]}), and an index counter (\textit{cidc}) that tracks the number of generated challenges for each chunk. The \textit{cidc} functions similarly to the \textit{idc}, ensuring the uniqueness of each chunk-level challenge.

In the PM-Dedup system, the initialization phase is defined by setting up two critical hash-maps: \textit{F} for files and \textit{C} for chunks. Each entry in these maps starts with uninitialized pointers, with \textit{ptr} in \textit{F} and \textit{cptr} in \textit{C} set to null. The arrays of precomputed challenge/response pairs (\textit{fileChal/fileRes[]}) in \textit{F} and (\textit{chunkChal/chunkRes[]}) in \textit{C} are initialized as empty arrays. Furthermore, the indexes \textit{idc} in \textit{F} and \textit{cidc} in \textit{C} are all set to zero.

\textbf{\textit{Challenge and Response Generation:}}

\begin{algorithm}[!t]
\caption{Generate Challenges}
\label{alg:challenge}
\begin{algorithmic}[1]
\STATE \textbf{Input:} $F, d, n, SMK$
\STATE \textbf{Output:} Modified $F$

\STATE \textbf{Function} GenChallenges($F, d, n, SMK$)
    \FOR{$i = 1$ \TO $n$}
        \STATE $F[d].idc \gets F[d].idc + 1$
        \STATE $s \gets$ GenSeed($SMK, d, F[d].idc$)
        \STATE $res \gets$ GenResp($s, F[d].ptr$)
        \STATE $F[d].fileRes \gets F[d].fileRes + [res]$
        \FORALL{$chunk$ in $F[d].chunks$}
            \STATE $cRes \gets$ GenResp($s, chunk.ptr$)
            \STATE $F[d].chunkRes[chunk.id] \gets F[d].chunkRes[chunk.id] + [cRes]$
        \ENDFOR
    \ENDFOR
    \RETURN $F$
\STATE \textbf{End Function}

\STATE \textbf{Function} GenSeed($SMK, d, idc$)
    \RETURN $SMK(d \parallel idc)$
\STATE \textbf{End Function}

\STATE \textbf{Function} GenResp($s, ptr$)
    \STATE SetSeed($s$)
    \STATE $res \gets$ ``
    \FOR{$j = 1$ \TO $K$}
        \STATE $pos \gets$ RandPos($ptr$)
        \STATE $res \gets res +$ BitAtPos($ptr, pos$)
    \ENDFOR
    \RETURN $res$
\STATE \textbf{End Function}
\end{algorithmic}
\end{algorithm}

The cloud server, recognizing the computational overhead and resource constraints of pre-computing challenges and responses for every chunk and file, selectively pre-computes these values only for those chunks in the share-index and local-index, and for files recorded in the local-index. Because these indexes are specifically optimized to enhance deduplication efficiency. By aligning the PoW process with these indexes, we ensure that the data verification process is similarly optimized.

As mentioned earlier, each unique file or chunk has a array of pre-computed chunk-level challenge/response pairs in the hash maps, which is used for PoW verification in subsequent requests. Therefore, during periodic updates of the pre-computed challenges and responses, the cloud server generates new challenges and responses for files or chunks that have been newly added to the share-index and local-index since the last update period, as well as for those where the pre-computed challenges and responses have been exhausted. For the file which needs to generate the challenge and response, uniquely identified by its hash \( \textit{d} \), as shown in Algorithm ~\ref{alg:challenge}, the cloud server is responsible for computing a predetermined number of challenges \( \textit{n} \). This computation starts by incrementing the \( \textit{idc} \) index, which ensures the uniqueness of each challenge. The next step involves generating a seed \( \textit{s} \). 
The purpose of generating the seed is to ensure that each challenge is unique and secure, as well as to facilitate cooperation with the client. It is created using the Cloud Server Master Key (CSMK), a secure secret key held by the cloud server, combined with the file's hash \(\textit{d}\) and the current challenge index \(\textit{idc}\) through a cryptographic function. The seed generation process can be represented as \(\textit{s} = \text{CSMK}(\textit{d} \parallel \textit{idc})\). This seed is then used to seed a pseudo-random number generator \( F_{\textit{s}} \) and the seed is stored in \( \textit{fileChal}[] \) as one of the \( \textit{n} \) file-level challenges.

With \( \textit{s} \) seeding \( F_{\textit{s}} \), the generator produces \( K \) random positions. The cloud server accesses the bit at each \( \textit{pos} \) in the file and appends it to an initially empty response string \( \textit{res} \). This string \( \textit{res} \), once fully formed, is stored in corresponding \( \textit{fileRes}[] \) for file-level responses. Thus, one of the \( \textit{n} \) pre-computed challenge and response pairs (\(\textit{fileChal/fileRes[]}\)) is generated. The remaining \( \textit{n-1} \) pairs are generated in the same manner. The same process is applied for generating chunk-level challenges and responses. The use of the CSMK in seed generation, combined with a seeded pseudo-random number generator, ensures that each challenge is not only unique but also secure, thereby making the fabrication of valid responses by unauthorized entities extremely difficult.

After generating the complete set of challenges and responses for each selected file and chunk, the cloud server allocates different subsets of these pre-computed pairs to various edge servers. This allocation ensures that each edge server receives a unique set of challenge/response pairs, derived from the larger pool generated by the cloud server. This approach prevents duplication of challenge/response pairs across edge servers, thereby maintaining the integrity and security of the PoW process while allowing each edge server to perform efficient and independent data verification.

\textbf{\textit{Verification:}}

\begin{algorithm}[!t]
\caption{Verify Ownership}
\label{alg:owner}
\begin{algorithmic}[1]
\STATE \textbf{Input:} $f, F, d$
\STATE \textbf{Output:} Boolean value indicating ownership verification

\STATE \textbf{Function} VerifyOwn($f, F, d$)
    \IF{$d \in F$}
        \STATE $s \gets F[d].fileRes[F[d].idu]$
        \STATE $cRes \gets$ ClientGenResp($f, s$)
        \IF{$cRes = F[d].fileRes[F[d].idu]$}
            \RETURN \textit{true}
        \ELSE
            \RETURN VerifyChunks($f, F, d$)
        \ENDIF
    \ELSE
        \RETURN \textit{false}
    \ENDIF
\STATE \textbf{End Function}

\STATE \textbf{Function} ClientGenResp($f, s$)
    \STATE SetSeed($s$)
    \STATE $res \gets$ ``
    \FOR{$j = 1$ \TO $K$}
        \STATE $pos \gets$ RandPos($f$)
        \STATE $res \gets res +$ BitAtPos($f, pos$)
    \ENDFOR
    \RETURN $res$
\STATE \textbf{End Function}

\STATE \textbf{Function} VerifyChunks($f, F, d$)
    \FORALL{$chunk$ in $F[d].chunks$}
        \STATE $s \gets F[d].chunkRes[chunk.id][F[d].idu]$
        \STATE $cRes \gets$ ClientGenResp($f, s$)
        \IF{$cRes \neq F[d].chunkRes[chunk.id][F[d].idu]$}
            \RETURN \textit{false}
        \ENDIF
    \ENDFOR
    \RETURN \textit{true}
\STATE \textbf{End Function}
\end{algorithmic}
\end{algorithm}
Initially, the client chunks the file and encrypts each chunk using keys obtained from the Key Server, generating fingerprints for each encrypted chunk. The client then uploads the hash of the entire file, represented as \( \textit{d} \), and the encrypted fingerprints of each chunk to the edge server. Upon receiving these hashes and fingerprints, the edge server follows a systematic verification process:

As shown in Algorithm ~\ref{alg:owner}, the edge server first checks if the file hash \( \textit{d} \) is present in its file-level hash-map \textit{F}. If \( \textit{d} \) is found, the edge server initiates file-level verification by selecting an unused file-level challenge from \textit{fileChal}[], sending the seed to the client, and comparing the client-generated response (derived from \( K \) random positions within the file) with the stored response in \textit{fileRes}[]. A match confirms the client's ownership of the entire file, allowing the edge server to proceed with the source-based deduplication process for the file.

If the file hash \( \textit{d} \) is not found in \textit{F}, the edge server proceeds to verify each chunk. The edge server checks the chunk-level hash-map \textit{C} for each chunk fingerprint. For chunk fingerprints found in \textit{C}, the edge server selects unused chunk-level challenges from \textit{chunkChal}[] and sends the seeds to the client. For chunk fingerprints not found in \textit{C}, the edge server uploads these chunk fingerprints to the cloud server to request the corresponding pre-computed challenges and responses. If the cloud server can generate challenges and responses for these chunks, it will send them back to the edge server. The edge server then selects the appropriate challenges and sends them to the client for ownership proof. The client uses the received seeds to generate responses by collecting bit-values at \( K \) random positions within the chunks. These responses are sent back to the edge server for verification. Successfully responding to these challenges demonstrates the client's ownership of the chunks. The edge server can then proceed with the source-based deduplication process for the verified chunks, which includes checking the local index first, then the share-index, and finally the cloud, as detailed in Section \ref{sec:over arc}. Only the unique chunks are uploaded to the cloud.

If the verification fails, the client is marked as suspicious. Repeated verification failures can lead to a temporary or permanent suspension of service to prevent potential abuse or security breaches.

Dual-Level lightweight and responsive PoW manages data verification by distributing pre-computed challenges and responses from the cloud server to the edge server, reducing latency. This system uses a dual-level verification method: file-level for quick checks and chunk-level for detailed scrutiny. This approach ensures thorough data validation while maintaining operational efficiency, making it suitable for large-scale systems.

\subsection{Edge Server Management} \label{sec:edge}
The edge server plays a critical role in the PM-Dedup system, acting as an intermediary between clients and the cloud server. It handles client requests for deduplication and ensures data integrity and security through the use of SGX enclaves.

\subsubsection{Enclave Management}

\paragraph{\textbf{Enclave Initialization and Secure Channel:}}

We follow established methods for the initial setup and secure communication of the enclave \cite{ren2021accelerating, kim2019shieldstore, wang2023svtpm, oh2020trustore}. The process begins with the establishment of an enclave on the edge server. This involves allocating a secure memory area, referred to as the Enclave Page Cache (EPC), and employing specialized SGX instructions to construct an isolated execution environment. During its creation, the contents of the enclave are cryptographically hashed, a critical step for verifying its authenticity in the subsequent attestation phase.

Following creation, the enclave initiates a remote attestation process to establish its trustworthiness. This procedure involves generating an attestation report containing the enclave's measurement, which the cloud server verifies to confirm the enclave's authenticity. Furthermore, the setup phase includes establishing a secure key exchange via the Elliptic Curve Diffie-Hellman (ECDH) algorithm. This key exchange between the enclave and the cloud server results in the derivation of a shared secret ($K_{\text{shared}}$), which is crucial for encrypting all subsequent communications between the two entities.

\paragraph{\textbf{Share-Index Management:}}

The edge server maintains a Share-Index, which is a selectively compiled list of fingerprints generated by the cloud server. This Share-Index is stored securely within the SGX enclave and is used to enhance the efficiency of source-based deduplication by ensuring high hit ratios during the duplicate check process. 

\textbf{\textit{Updating the Share-Index:}} The Share-Index is updated when the edge server observes that its hit ratio drops below a certain threshold after a duration. Upon detecting this, the edge server sends a request to the cloud server for an updated Share-Index. The cloud server then compiles a new Share-Index and transmits this updated list to the edge server, as discussed in detail in Section 4.1.

\textbf{\textit{Transmission of Updates:}} Updates to the Share-Index are transmitted from the cloud server to the edge server through a secure channel established during the enclave initialization phase. These updates are encrypted using the shared secret ($K_{\text{shared}}$) to ensure confidentiality during transit. Once received, the SGX enclave decrypts the updates and verifies their integrity before incorporating them into the existing Share-Index.

\textbf{\textit{Deletion of Obsolete Entries:}} As part of the update process, the edge server also removes obsolete fingerprints from the Share-Index. This involves identifying entries that are no longer relevant or less frequently accessed and hen securely delete them from the SGX enclave to maintain an efficient and up-to-date Share-Index.

\paragraph{\textbf{Management of Pre-computed Challenges and Responses:}}

The management of pre-computed challenges and responses in the edge server's SGX enclave relies on the share index and local index. Updates and deletions in file-level and chunk-level hashmaps are triggered by changes in these indexes. During periodic updates, the cloud server generates new chunk-level and file-level precomputed challenges and responses for files or chunks that have been added since the last update or where pairs have been exhausted. Afterward, the cloud server sends the updated pre-computed challenges and responses to the edge server. Upon sharing these pre-computed values with the edge server, the cloud server marks the shared portions of both file-level and chunk-level pre-computed challenges and responses as invalid in the cloud-level hash map to ensure security. 

\subsubsection{Local Index Management}


The local-index reflects the historical uploads from clients within the same branch and includes both chunk and file fingerprints. For each type of fingerprint, it is implemented using a hash map to store the fingerprints and a separate doubly linked list tracks the usage order for implementing the LRU strategy.

The local-index is initialized with a fixed capacity based on the edge server's storage and memory constraints. When a new chunk or file is uploaded by a client, the edge server checks if the fingerprint is already present in the local-index. If the fingerprint is found, the corresponding entry in the linked list is moved to the front to mark it as recently used. If the fingerprint is not present, it is added to the hash map, and a new entry is inserted at the front of the linked list. If the local-index has reached its capacity, the least recently used entry is removed from both the hash map and the linked list to make space for the new entry, ensuring that the local-index remains up-to-date with the most recently accessed data.

\section{Implementation}\label{sec:implementation}

The PM-Dedup system, consisting approximately 3.2 K lines of C code, incorporates OpenSSL 1.1.1 for cryptographic processes and Intel SGX SDK 2.7 to manage secure enclaves. It utilizes SHA-256 for creating fingerprints of data chunks and AES-256 for the encryption of sensitive data.

On the client side, FastCDC \cite{xia2016fastcdc} is implemented for content-defined chunking. The edge server setup includes a Diffie-Hellman Key Exchange (DHKE) mechanism based on the NIST P-256 elliptic curve, facilitating secure communications \cite{ren2021accelerating}. For example, when the cloud handles 20 gigabytes (GB) of data uploaded from the client with a chunk size of 16 kilobytes (KB), and the share-index covers the 10\% of the data, the edge server stores metadata including the share-index, precomputed challenge-response pairs, and other relevant data. These metadata are updated after every 1 GB of uploads. Under these conditions, the metadata will take up approximately 80 megabytes (MB) of the SGX enclave. Note that these values can be adjusted based on chunk size, enclave capacity, the number of precomputed challenge-response pairs per chunk/file, and other factors.

We chose not to fix the size of the local index, and the size of data that can be stored in the SGX enclave, because doing so could lead to disproportionate results between experiments with varying data sizes. For instance, if the sizes were fixed, smaller datasets might result in nearly all relevant metadata being stored on the edge server, making it difficult to assess the effectiveness of the metadata selection.  To maintain consistency and fairness in our evaluations, we scale the size of the local-index and the share-index relative to the total data volume in the cloud, but without exceeding the physical limits of the edge server’s resources.

The prototype of the PM-Dedup system is implemented on 5 Dell PowerEdge R430 servers, with 4 machines designated for clients and one for the cloud server components. Each machine is equipped with a Seagate ST1000NM0033-9ZM173 SATA hard disk, offering a storage capacity of 1TB. These machines are powered by a 2.40GHz Intel Xeon processor with 24 cores and supported by 64GB of RAM. The edge server prototype is set up on a machine with a 3.2GHz Intel Core i7-7700K CPU, 32 GB of RAM, and a 1TB Samsung 980 PRO PCIe 4.0 NVMe SSD. All machines in this setup run on Ubuntu 16.04.

\section{Evaluation}\label{sec:evaluation}

In this paper, we utilize four widely used real-world backup datasets and one self-collected dataset to conduct our observations in Sections ~\ref{sec:Real-World_Latency} and ~\ref{sec:shine}, and our evaluations in Sections ~\ref{sec:6.2} and ~\ref{sec:Share-index Selection Analysis}, as detailed in Table ~\ref{tab:deduplication_summary}.

\begin{itemize}
    \item \textbf{MS:} This dataset contains 30 Windows snapshots ~\cite{meyer2012study}.
    \item \textbf{UBUNTU:} This dataset contains 12 consecutive Ubuntu snapshots ~\cite{ubuntu}.
    \item \textbf{FSL:} This dataset contains 20 consecutive homes snapshots ~\cite{sun2016long}, which are collected by the File system and Storage Lab (FSL) at Stony Brook University.
    \item \textbf{LAB:} This dataset comprises 33 snapshots of lab servers, collected by our team.
    \item \textbf{GCC:} This dataset contains 24 non-consecutive versions of unpacked source code ~\cite{cheng2021coupling}.
\end{itemize}

For convenience and focus, we simulated the performance within a single branch setup. Four client machines represented the clients within a branch, with one edge server and one cloud server. To emulate different branches within an organization, we pre-uploaded various snapshot versions to the cloud server, thereby creating a baseline of common data that different branches would typically share. Following this, we uploaded different snapshot versions to test the system’s efficiency and performance in handling deduplication and ownership verification.

\begin{table}[!t]
\centering
\caption{Characteristics of Real-world Datasets}
\label{tab:deduplication_summary}
\resizebox{0.49\textwidth}{!}{
\begin{tabular}{|c|c|c|c|}
\hline
\textbf{Dataset} & \textbf{Snapshots} & \textbf{Size} & \textbf{Deduplication Ratio} \\
\hline
FSL & 20 & 2.1 TB & 11.8 \\
\hline
MS & 30 & 3.6 TB & 5.1 \\
\hline
LAB & 33 & 15.3 GiB & 27.1 \\
\hline

UBUNTU & 12 & 60.3 GiB & 4.1 \\
\hline

GCC & 24 & 7.3 GiB & 1.4 \\
\hline
\end{tabular}
}
\end{table}

\subsection{Real-World Latency Analysis} \label{sec:Real-World_Latency}

The latency of clients when accessing edge servers compared to cloud servers can differ significantly, influenced by factors such as geographic location and the choice of cloud provider, which can introduce variability and affect the accuracy of our experiments. To address this, we conducted an analysis based on the real-world dataset reported by Charyyev et al.~\cite{charyyev2020latency}, which assessed latency from 8,456 end-users to 6,341 edge servers and 69 cloud locations. By conducting this analysis, we aim to quantify the latency improvements that edge servers can provide, and use these quantified improvements as a baseline for our subsequent experiments.

\begin{figure}[!t]
  \centering
  \includegraphics[width=\columnwidth]{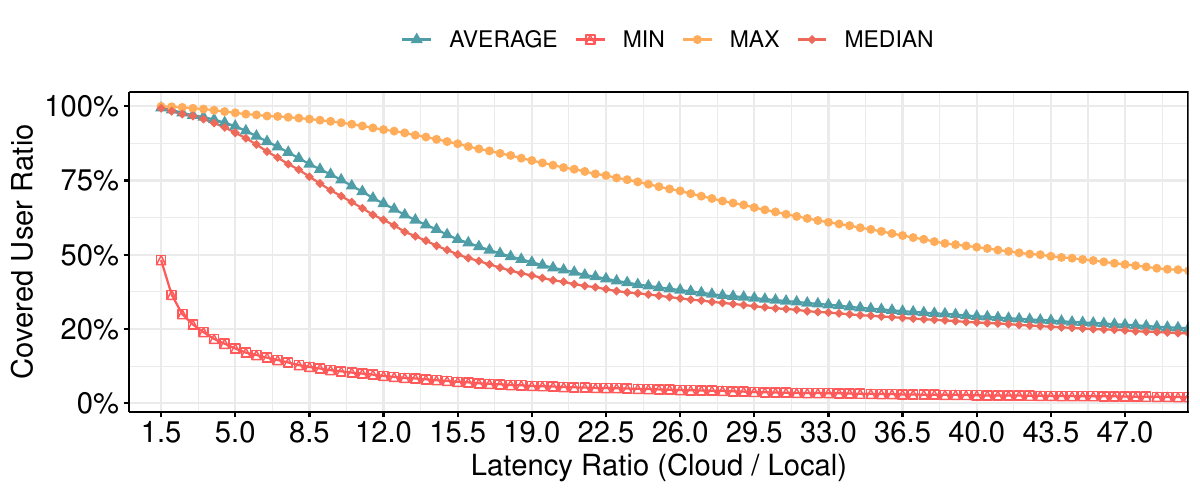} 
  \caption{Latency Ratio vs. User Coverage}
  \label{fig:user cover}
\end{figure}

For each user's latency in accessing the 69 cloud locations, we selected the minimum, maximum, median, and average values as our metrics. These values were then divided by the latency for accessing the edge server to calculate the latency ratio, which is represented on the x-axis in Figure ~\ref{fig:user cover} of our analysis. The covered user ratio, indicated on the y-axis, represents the proportion of users exceeding this latency ratio out of the total user count. The Figure ~\ref{fig:user cover} illustrates that, for the average, median, and maximum metrics, over 75\% of users experience a latency ratio exceeding 8.5, indicating that the time taken to access cloud services is 8.5 times greater than that required to access edge servers. Additionally, even for the minimum metrics, over 50\% of users still experience a latency ratio exceeding 1.5.

To gain further insights, we analyzed the average and median values of the four types of latency ratios, with results presented in Table ~\ref{tab:memory_overhead}. For the average, median, and maximum latency ratios, both the average and median values of them exceed 15.5. Notably, even for the minimum latency ratio, its average value stands at 6.89, indicating that, on average, the fastest cloud access time is still 6.89 times longer than accessing edge servers. We used this ratio as the baseline for our subsequent experiments to highlight the latency reduction benefits of edge server utilization.


\begin{table}[!t]
\centering
\caption{Average and Median Latency Ratios}
\label{tab:memory_overhead}
\resizebox{0.49\textwidth}{!}{
\begin{tabular}{|c|c|c|}
\hline
\textbf{Metric} & \textbf{Average Latency Ratio} & \textbf{Median Latency Ratio} \\
\hline
AVERAGE & 50.96 & 17.79 \\
\hline
MEDIAN & 45.98 & 15.59 \\
\hline
MIN & 6.89 & 1.67 \\
\hline
MAX & 119.72 & 43.01 \\
\hline
\end{tabular}
}
\end{table}



\subsection{Latency Reduction Measurement}\label{sec:6.2}

This experiment aims to assess the latency reduction performance of PM-Dedup. We compare PM-Dedup with DupLESS \cite{keelveedhi2013dupless}, a classical secure deduplication approach that employs a server-aided architecture to thwart offline brute force attacks without using a secure zone. Another comparison is made with SGXDedup \cite{ren2021accelerating}, which utilizes Intel SGX on both the client and key server sides to accelerate encrypted deduplication. Our comparisons are based on the conceptual frameworks and methodologies outlined in these approaches. Additionally, we conduct a comparison within PM-Dedup itself, testing the performance of the share-index by excluding the local-index to evaluate the effectiveness of the cross-branch share-index alone.

For an intuitive presentation of the results, we analyze the latency involved in uploading a snapshot using all four approaches: PM-Dedup, DupLESS, SGXDedup, and PM-Dedup without lcoal-index. For accuracy, our experiments were performed on four datasets. For each dataset, we executed the experiments 10 times. Each time, we randomly selected several snapshots pre-uploaded to the cloud and uploaded a different snapshot to measure its latency. We standardized all latency results to the equivalent of 1 GB of data upload for consistency. The following sections will analyze how PM-Dedup optimizes each stage of the data upload process to improve overall latency reduction performance.

\textbf{\textit{Exp 1: Overall latency reduction:}}

 \begin{figure}[!t]
  \centering
  \includegraphics[width=\columnwidth]{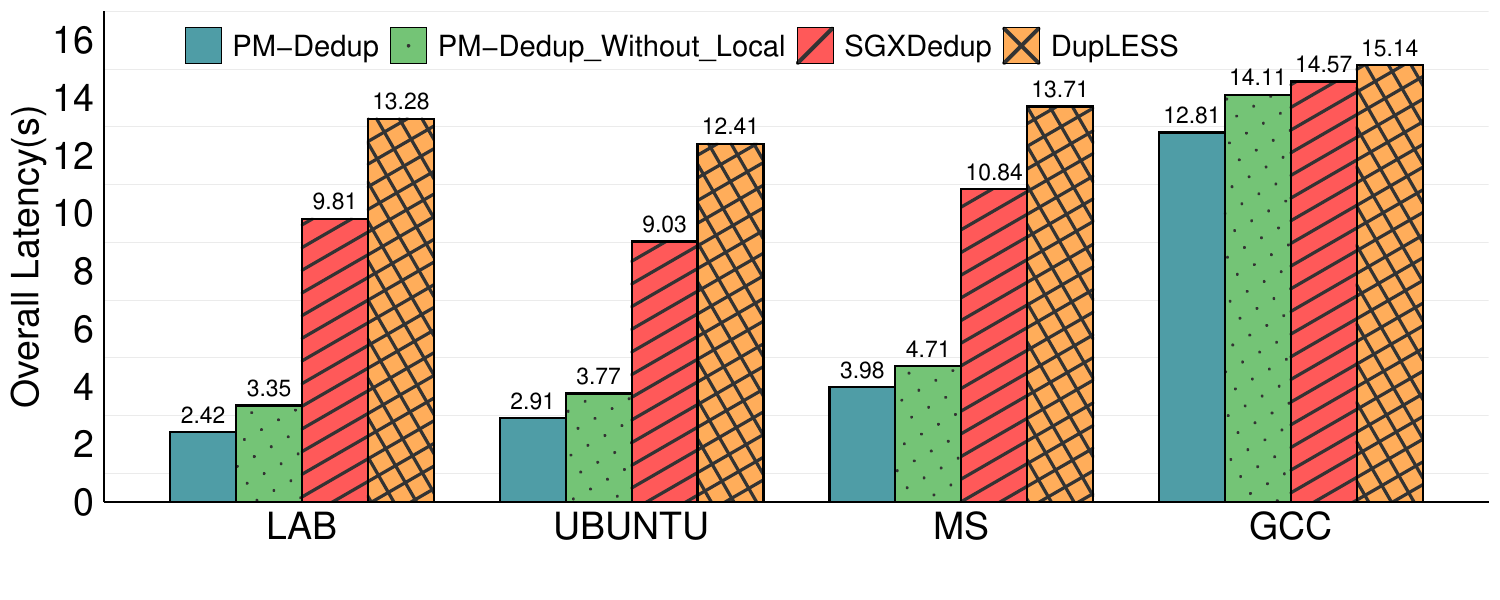} 
  \caption{Overall latency comparison}
  \label{fig:overall}
\end{figure}

 This section presents an analysis of the overall latency, which is defined as the interval from data upload by the client to the final acknowledgment from the cloud, including all intermediate steps such as deduplication checks, proof-of-ownership verification, data transfer, among others.

As shown in Figure ~\ref{fig:overall}, the overall latency of our prototype is closely related to data redundancy. Our model performs best when data redundancy is high, achieving an 81.8\% reduction in latency compared to DupLESS in the LAB dataset. This significant improvement is due to the effectiveness of PM-Dedup in reducing cloud interactions by migrating a portion of the PoW and deduplication check tasks to the edge server. This allows for quicker verification and deduplication processes at the edge, minimizing the need for high-latency cloud queries.

However, in datasets with lower redundancy, such as the GCC dataset, the latency reduction is less pronounced. PM-Dedup still outperforms SGXDedup and DupLESS, albeit by a smaller margin. In low redundancy scenarios, the edge server cannot effectively perform deduplication checks due to the lack of duplicate chunks, meaning that a significant portion of the checks still need to be escalated to the cloud server. Consequently, the additional processing at the edge server can sometimes offset the latency gains.

When comparing PM-Dedup with PM-Dedup without the local index, the inclusion of the local index significantly reduces latency. In the LAB dataset, PM-Dedup outperforms by up to 30\%. This underscores the importance of utilizing both local and shared indexes for optimal performance.

The following sections will analyze how our prototype optimizes each stage of the data upload process to improve overall latency reduction performance.

 \textbf{\textit{Exp 2: Ownership verification latency reduction:}}

 \begin{figure}[!t]
  \centering
  \includegraphics[width=\columnwidth]{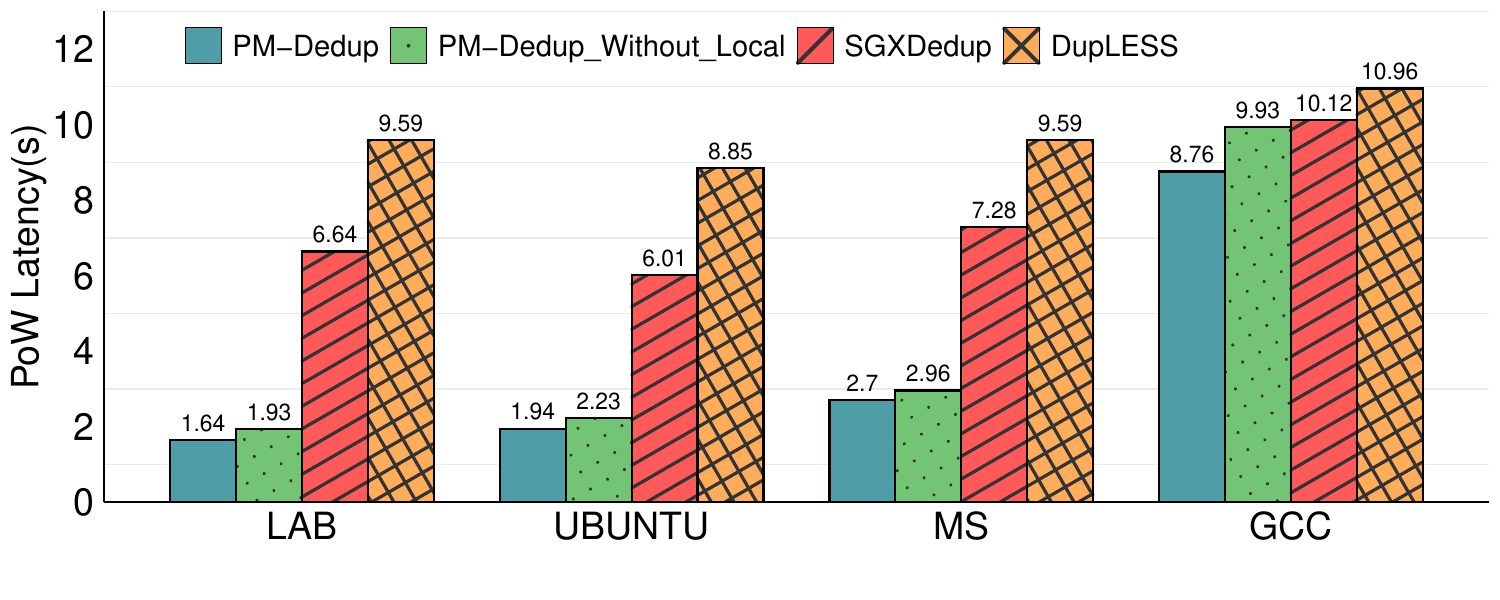} 
  \caption{Ownership verfication latency}
  \label{fig:pow}
\end{figure}

In the PoW latency evaluation, we measure the ownership verfication latency, including the challenge-response protocol at both the edge and cloud servers. The evaluation focuses on the average latency per chunk across the dataset uploaded by the client. As shown in Figure ~\ref{fig:pow}, PM-Dedup significantly reduces latency compared to SGXDedup and DupLESS, especially in high redundancy datasets.

For example, in the LAB dataset, PM-Dedup achieves a 75.3\% latency reduction compared to SGXDedup and an 82.9\% reduction compared to DupLESS. This improvement is attributed to our lightweight PoW mechanism and the offloading of PoW tasks to the edge server, which reduces the high latency associated with cloud-level checks.

When comparing PM-Dedup to PM-Dedup without local-index, the latencies are similar because PoW operations require substantial time for challenge-response verification regardless of the local index. The local index mainly improves deduplication check efficiency but has less impact on PoW latency as both models perform comprehensive PoW procedures.

 \textbf{\textit{Exp 3: Deduplication Check latency reduction:}}

 \begin{figure}[!t]
  \centering
  \includegraphics[width=\columnwidth]{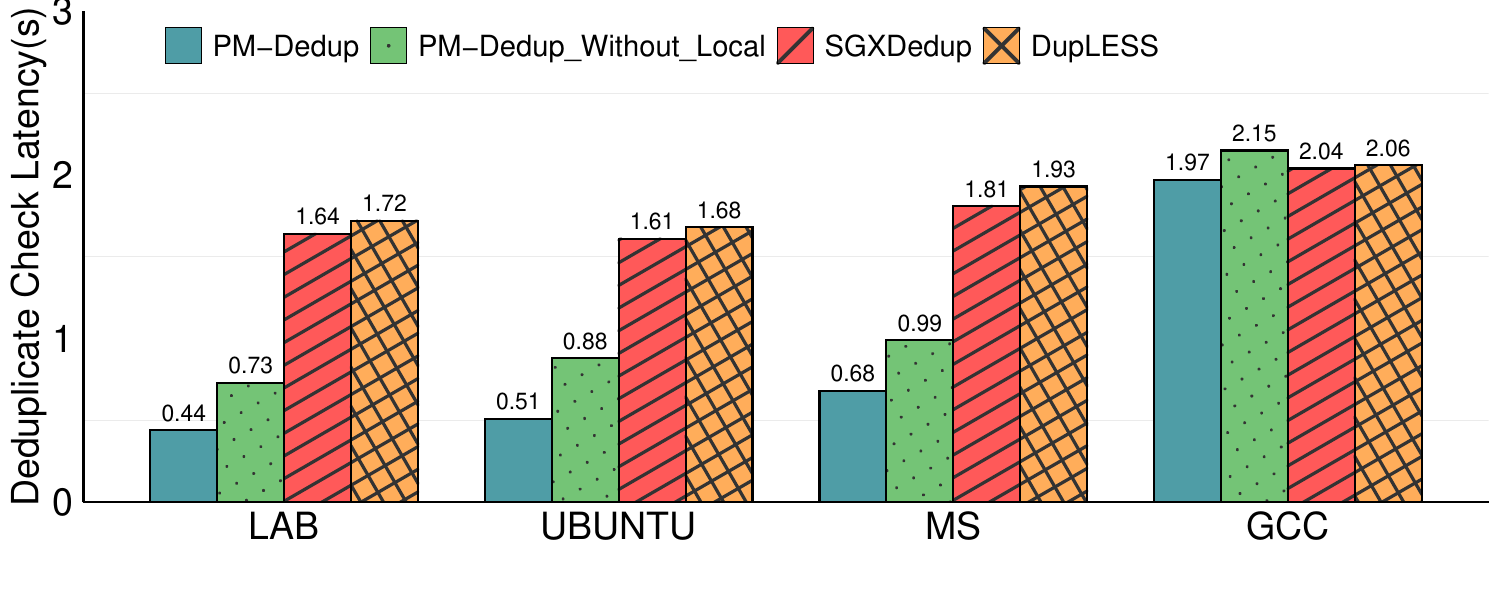} 
  \caption{Deduplication check latency}
  \label{fig:check}
\end{figure}

We measure the deduplication check latency, which includes the time taken for share-index selection, edge server-level check, and cloud-level check. Our model exhibits optimal performance when data redundancy is high. As shown in Figure ~\ref{fig:check}, in the LAB dataset, our prototype achieves a latency reduction of 74.4\% compared to DupLESS and 73.2\% compared to SGXDedup. This significant improvement is due to our model's ability to transfer many deduplication checks to the edge server, minimizing the need for high-latency cloud-level checks.

When comparing PM-Dedup with PM-Dedup without the local-index, we observe a considerable performance difference. For example, in the LAB dataset, PM-Dedup achieves a 39.7\% lower latency. This reduction in latency is attributed to the presence of the local-index, which reduces the need for frequent and costly SGX enclave operations, emphasizing the importance of utilizing both local and share-indexes for optimal performance.

\begin{figure}[!t]
  \centering
  \includegraphics[width=0.9\columnwidth]{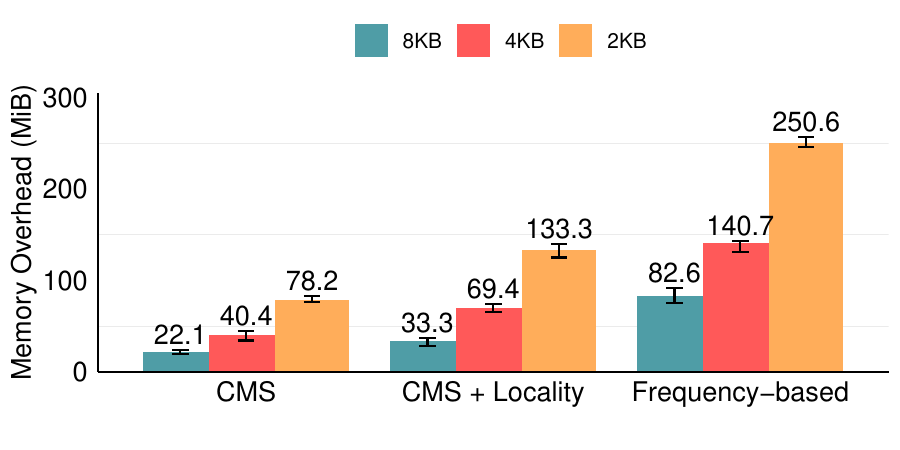} 
  \caption{Memory overhead}
  \label{fig:memory}
\end{figure}

\begin{figure}[!t]
  \centering
  \includegraphics[width=0.9\columnwidth]{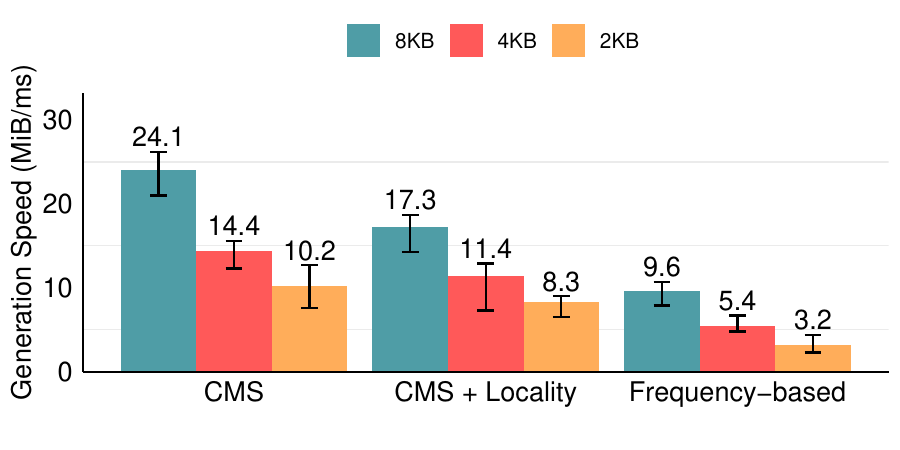} 
  \caption{Share-index generation speed}
  \label{fig:speed}
\end{figure}

\begin{figure*}[!t]
  \centering
  \begin{minipage}[b]{\columnwidth}
    \centering
    \includegraphics[width=\linewidth]{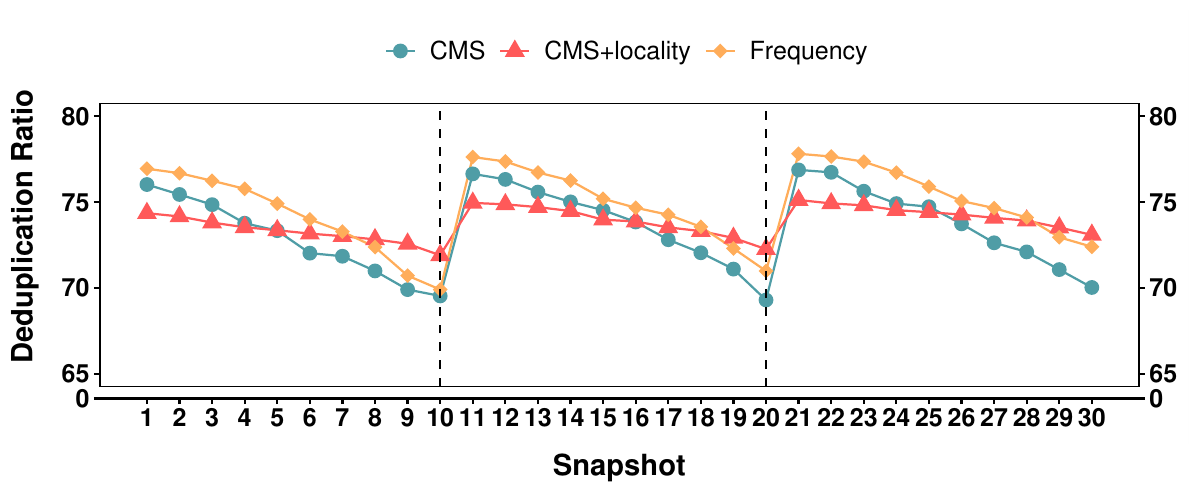}
    \caption{MS Dataset}
    \label{fig:ex1}
  \end{minipage}
  \hfill
  \begin{minipage}[b]{\columnwidth}
    \centering
    \includegraphics[width=\linewidth]{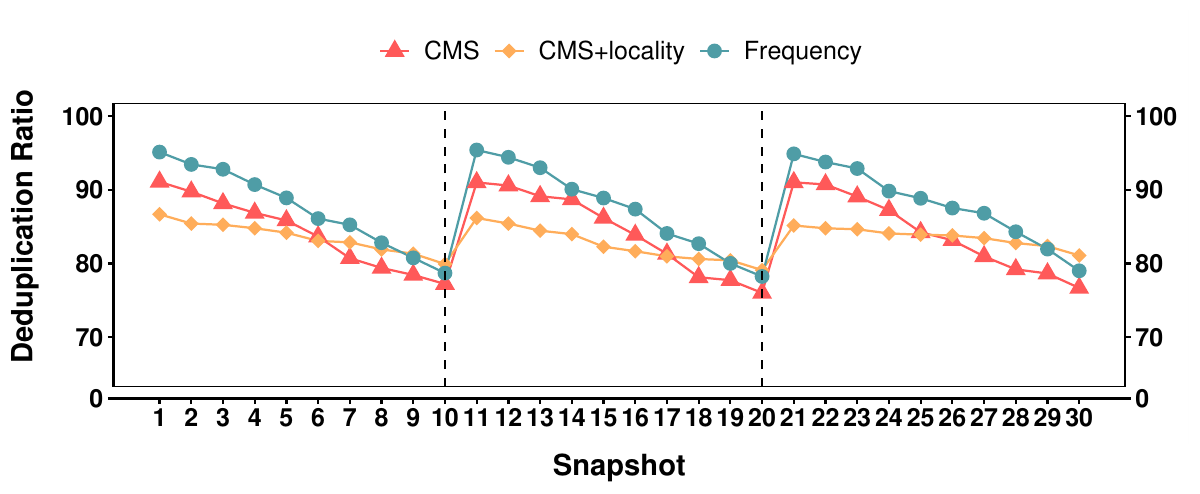}
    \caption{LAB Dataset}
    \label{fig:ex2}
  \end{minipage}
  
\end{figure*}

\subsection{Share-index Selection Analysis} \label{sec:Share-index Selection Analysis}

Share-index selection significantly impacts deduplication efficiency. We compare our two share-index selection approaches with the most commonly used approach called frequency-based selection. This approach involves maintaining a counter for each chunk to track its appearance time, thereby selecting the most frequent subset of all chunks based on this appearance time. The objective of these experiments is to comprehensively measure and evaluate the three schemes, with a focus on deduplication ratio, CPU utilization, and storage space saving. For the locality plus CMS share-index selection, we configure it such that 90\% of the chunks are selected via CMS selection, while the remaining are chosen based on logical locality selection. Importantly, the number of fingerprints selected by each of the three methods is kept consistent to ensure a fair comparison.


\textbf{\textit{Memory Overhead Measurement:}}
We analyzed the memory overhead associated with processing data chunks of 2 KB, 4 KB, and 8 KB sizes, presenting the overhead for every 100 GB of data processed. Our analysis was based on approximately 800 GB of MS snapshots. Note that, to capture the additional memory overhead, we monitored the size of the data structures maintained by each approach during the program's execution.

Taking the 2 KB chunking as an example, Figure ~\ref{fig:memory} shows that CMS selection has an average memory overhead of 22.1 MB, which is 73.2\% lower than the 82.6 MB overhead of frequency-based methods. However, when CMS incorporates locality, the overhead slightly increases to 33.3 MB due to the additional space required for locality measurement, yet it still maintains a 59.7\% lower overhead compared to frequency-based approaches.

\textbf{\textit{Share-index Generation Speed Measurement:}}
We investigated the generation speed of share-indexes for processing data chunks with sizes of 2 KB, 4 KB, and 8 KB. Figure 8 shows that the CMS approach yields the highest generation speed, followed by CMS combined with locality, and lastly, the frequency-based selection. Taking 2 KB chunk processing as an example, as shown in Figure ~\ref{fig:speed}, CMS approach is approximately 3.19 times faster than the frequency-based approach and the CMS combined with locality is about 2.59 times faster than the frequency-based approach. The superior performance of the CMS approach is attributed to the frequency-based method requiring frequent counter updates, which leads to additional time overhead. The CMS combined with locality is slightly less efficient than the pure CMS approach due to the time-consuming of locality detection.

\textbf{\textit{Deduplication Ratio Measurement:}} We conducted two experiments, one with 30 consecutive MS snapshots and another with 30 consecutive LAB snapshots, to assess the deduplication ratio. Before beginning the tests, we select the share-index from previous snapshots using three selection approaches and compare this share-index with consecutive snapshots following these snapshots. The deduplication ratio on the y-axis represents the total size of duplicate chunks post-comparison divided by the total size of the snapshot. In Figure ~\ref{fig:ex1} and Figure ~\ref{fig:ex2}, the x-axis represents snapshots, with each point referring to the consecutive snapshot following the previous one. Note that, we did not update the share-index immediately after comparison; instead, we update it at snapshots 10 and 20 to investigate the long-term stability of these approaches.

Figure ~\ref{fig:ex1} and Figure ~\ref{fig:ex2} show the variation in the deduplication ratio with increasing snapshot numbers. The deduplication ratios for all three approaches initially decrease before updating the share-index, then significantly recover upon updating the share-index. This is because chunks that frequently appear may be absent in future snapshots. The frequency-based approach generally outperforms the other two, as it accurately selects the most frequent subset of all chunks. In contrast, the CMS selection is a probabilistic approach, where chunk selection is based on estimated frequencies rather than exact counts, making it a less precise choice for identifying the most frequent chunks.
In early snapshots, the CMS-based and frequency-based share-index selection shows superior performance compared to CMS plus locality, but its deduplication ratio declines over time. In contrast, although the CMS plus locality approach initially exhibits a lower deduplication ratio at the beginning, it achieves comparable or superior performance in later snapshots. This trend is attributed to the CMS plus locality approach incorporating a small proportion of logically related chunks that may not be present in the initial snapshots but appear in later ones, resulting in a slower rate of decline.

The experimental results show that frequency-based selection yields a superior deduplication ratio; however, it incurs significantly higher memory usage and computational costs. The other two approaches approximate the deduplication ratio of the frequency-based selection approach while consuming less memory and computational resources. If the share-index is frequently updated, the CMS selection is preferred due to its initial high deduplication ratio and lower resource consumption. Conversely, if the share-index is updated infrequently, combining CMS with locality-based selection yields a superior long-term performance.

\subsection{Where Do PM-Dedup Shine?} \label{sec:shine}

We explore scenarios where PM-Dedup are particularly effective. We conduct an analysis on five real-world data traces. For each trace, we evaluate the elimination ratio achieved by storing the metadata of top 5\%, 10\%, and 20\% of high-frequency chunks on the edge server. The elimination ratio, defined as the percentage of cloud-level deduplication checks that can be avoided by leveraging high-frequency chunks stored at the edge server. This ratio is calculated by dividing the total size of avoidable chunks by the overall size of chunks within each trace. 

\begin{figure}[!t]
  \centering
  \includegraphics[width=\columnwidth]{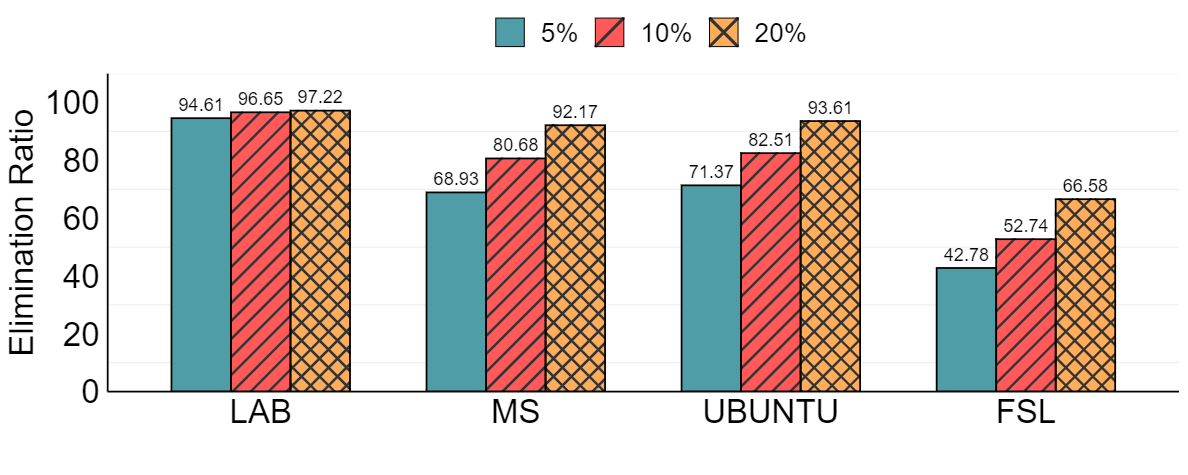} 
  \caption{Elimination ratio analysis} 
  \label{fig:elim}
  \begin{footnotesize}
    \begin{flushleft} 
    \end{flushleft}
  \end{footnotesize}
\end{figure}

Our observations shown in Figure ~\ref{fig:elim} indicate that PM-Dedup are well-suited for backup data, where redundancy is prevalent. For example, in the LAB trace, 5\% top high-frequency fingerprints can bring 94.61\% elimination ratio. However, for streams with predominantly new data at each instance, the use of edge servers for deduplication checks might not only be less beneficial but could also introduce inefficiencies, as shown in Section ~\ref{sec:6.2}. This is because the edge server adds an additional checkpoint that, when failing to complete PoW tasks or deduplication checks, results in unnecessary latency before resorting to cloud to finish the task. 
Therefore, PM-Dedup are better suited for data with high redundancy and less appropriate for scenarios with a high volume of new data.

\section{Conclusion}\label{sec:conclusion}

In this paper, we present PM-Dedup, which addresses the inherent challenges of security concern, high latency, and computational overhead in secure deduplication by distributing certain tasks from the cloud to edge servers. By employing efficient share-index generation and a dual-level lightweight and responsive PoW scheme, PM-Dedup enables the vast majority of secure deduplication tasks to be completed at the edge with low latency. The trusted zone is employed to safeguard shared data from the cloud, ensuring the efficiency of deduplication processes without sacrificing security. Our evaluations on real-world datasets demonstrate that PM-Dedup significantly reduces overall latency and improves data verification efficiency without compromising security.

\bibliographystyle{plain}
\bibliography{ref.bib}

\begin{thebibliography}{10}

\bibitem{ubuntu}
{Ubuntu Releases}.
\newblock Available: \url{https://old-releases.ubuntu.com/releases/}, 2024.
\newblock Accessed: 2024-02-26.

\bibitem{bellare2013message}
Mihir Bellare, Sriram Keelveedhi, and Thomas Ristenpart.
\newblock Message-locked encryption and secure deduplication.
\newblock In {\em Annual international conference on the theory and applications of cryptographic techniques}, pages 296--312. Springer, 2013.

\bibitem{chai2012verifiable}
Qi~Chai and Guang Gong.
\newblock Verifiable symmetric searchable encryption for semi-honest-but-curious cloud servers.
\newblock In {\em 2012 IEEE international conference on communications (ICC)}, pages 917--922. IEEE, 2012.

\bibitem{charyyev2020latency}
Batyr Charyyev, Engin Arslan, and Mehmet~Hadi Gunes.
\newblock Latency comparison of cloud datacenters and edge servers.
\newblock In {\em GLOBECOM 2020-2020 IEEE Global Communications Conference}, pages 1--6. IEEE, 2020.

\bibitem{cheng2021coupling}
Liangfeng Cheng, Yuchong Hu, Zhaokang Ke, and Zhongjie Wu.
\newblock Coupling right-provisioned cold storage data centers with deduplication.
\newblock In {\em Proceedings of the 50th International Conference on Parallel Processing}, pages 1--11, 2021.

\bibitem{cormode2005improved}
Graham Cormode and Shan Muthukrishnan.
\newblock An improved data stream summary: the count-min sketch and its applications.
\newblock {\em Journal of Algorithms}, 55(1):58--75, 2005.

\bibitem{di2012boosting}
Roberto Di~Pietro and Alessandro Sorniotti.
\newblock Boosting efficiency and security in proof of ownership for deduplication.
\newblock In {\em Proceedings of the 7th ACM symposium on information, computer and communications security}, pages 81--82, 2012.

\bibitem{di2016proof}
Roberto Di~Pietro and Alessandro Sorniotti.
\newblock Proof of ownership for deduplication systems: a secure, scalable, and efficient solution.
\newblock {\em Computer Communications}, 82:71--82, 2016.

\bibitem{duan2014distributed}
Yitao Duan.
\newblock Distributed key generation for encrypted deduplication: Achieving the strongest privacy.
\newblock In {\em Proceedings of the 6th edition of the ACM Workshop on Cloud Computing Security}, pages 57--68, 2014.

\bibitem{halevi2011proofs}
Shai Halevi, Danny Harnik, Benny Pinkas, and Alexandra Shulman-Peleg.
\newblock Proofs of ownership in remote storage systems.
\newblock In {\em Proceedings of the 18th ACM conference on Computer and communications security}, pages 491--500, 2011.

\bibitem{harnik2010side}
Danny Harnik, Benny Pinkas, and Alexandra Shulman-Peleg.
\newblock Side channels in cloud services: Deduplication in cloud storage.
\newblock {\em IEEE Security \& Privacy}, 8(6):40--47, 2010.

\bibitem{harnik2018securing}
Danny Harnik, Eliad Tsfadia, Doron Chen, and Ronen Kat.
\newblock Securing the storage data path with sgx enclaves.
\newblock {\em arXiv preprint arXiv:1806.10883}, 2018.

\bibitem{hasan2005evolution}
Ragib Hasan, William Yurcik, and Suvda Myagmar.
\newblock The evolution of storage service providers: techniques and challenges to outsourcing storage.
\newblock In {\em Proceedings of the 2005 ACM workshop on Storage Security and Survivability}, pages 1--8, 2005.

\bibitem{jauernig2020trusted}
Patrick Jauernig, Ahmad-Reza Sadeghi, and Emmanuel Stapf.
\newblock Trusted execution environments: properties, applications, and challenges.
\newblock {\em IEEE Security \& Privacy}, 18(2):56--60, 2020.

\bibitem{kamara2010cryptographic}
Seny Kamara and Kristin Lauter.
\newblock Cryptographic cloud storage.
\newblock In {\em International Conference on Financial Cryptography and Data Security}, pages 136--149. Springer, 2010.

\bibitem{kang2021iceclave}
Luyi Kang, Yuqi Xue, Weiwei Jia, Xiaohao Wang, Jongryool Kim, Changhwan Youn, Myeong~Joon Kang, Hyung~Jin Lim, Bruce Jacob, and Jian Huang.
\newblock Iceclave: A trusted execution environment for in-storage computing.
\newblock In {\em MICRO-54: 54th Annual IEEE/ACM International Symposium on Microarchitecture}, pages 199--211, 2021.

\bibitem{keelveedhi2013dupless}
Sriram Keelveedhi, Mihir Bellare, and Thomas Ristenpart.
\newblock $\{$DupLESS$\}$:$\{$Server-Aided$\}$ encryption for deduplicated storage.
\newblock In {\em 22nd USENIX security symposium (USENIX security 13)}, pages 179--194, 2013.

\bibitem{kim2019shieldstore}
Taehoon Kim, Joongun Park, Jaewook Woo, Seungheun Jeon, and Jaehyuk Huh.
\newblock Shieldstore: Shielded in-memory key-value storage with sgx.
\newblock In {\em Proceedings of the Fourteenth EuroSys Conference 2019}, pages 1--15, 2019.

\bibitem{kotla2007safestore}
Ramakrishna Kotla, Lorenzo Alvisi, and Mike Dahlin.
\newblock Safestore: A durable and practical storage system.
\newblock In {\em USENIX Annual Technical Conference}, pages 129--142, 2007.

\bibitem{luo2021resource}
Quyuan Luo, Shihong Hu, Changle Li, Guanghui Li, and Weisong Shi.
\newblock Resource scheduling in edge computing: A survey.
\newblock {\em IEEE Communications Surveys \& Tutorials}, 23(4):2131--2165, 2021.

\bibitem{malekzadeh2021honest}
Mohammad Malekzadeh, Anastasia Borovykh, and Deniz G{\"u}nd{\"u}z.
\newblock Honest-but-curious nets: Sensitive attributes of private inputs can be secretly coded into the classifiers' outputs.
\newblock In {\em Proceedings of the 2021 ACM SIGSAC Conference on Computer and Communications Security}, pages 825--844, 2021.

\bibitem{meyer2012study}
Dutch~T Meyer and William~J Bolosky.
\newblock A study of practical deduplication.
\newblock {\em ACM Transactions on Storage (ToS)}, 7(4):1--20, 2012.

\bibitem{miao2015secure}
Meixia Miao, Jianfeng Wang, Hui Li, and Xiaofeng Chen.
\newblock Secure multi-server-aided data deduplication in cloud computing.
\newblock {\em Pervasive and Mobile Computing}, 24:129--137, 2015.

\bibitem{naor2004number}
Moni Naor and Omer Reingold.
\newblock Number-theoretic constructions of efficient pseudo-random functions.
\newblock {\em Journal of the ACM (JACM)}, 51(2):231--262, 2004.

\bibitem{oh2020trustore}
Hyunyoung Oh, Adil Ahmad, Seonghyun Park, Byoungyoung Lee, and Yunheung Paek.
\newblock Trustore: Side-channel resistant storage for sgx using intel hybrid cpu-fpga.
\newblock In {\em Proceedings of the 2020 ACM SIGSAC Conference on Computer and Communications Security}, pages 1903--1918, 2020.

\bibitem{ren2021accelerating}
Yanjing Ren, Jingwei Li, Zuoru Yang, Patrick~PC Lee, and Xiaosong Zhang.
\newblock Accelerating encrypted deduplication via $\{$SGX$\}$.
\newblock In {\em 2021 USENIX Annual Technical Conference (USENIX ATC 21)}, pages 957--971, 2021.

\bibitem{sabt2015trusted}
Mohamed Sabt, Mohammed Achemlal, and Abdelmadjid Bouabdallah.
\newblock Trusted execution environment: What it is, and what it is not.
\newblock In {\em 2015 IEEE Trustcom/BigDataSE/Ispa}, volume~1, pages 57--64. IEEE, 2015.

\bibitem{shamma2011capo}
Mohammad Shamma, Dutch~T Meyer, Jake Wires, Maria Ivanova, Norman~C Hutchinson, and Andrew Warfield.
\newblock Capo: Recapitulating storage for virtual desktops.
\newblock In {\em 9th USENIX Conference on File and Storage Technologies (FAST 11)}, 2011.

\bibitem{shin2017survey}
Youngjoo Shin, Dongyoung Koo, and Junbeom Hur.
\newblock A survey of secure data deduplication schemes for cloud storage systems.
\newblock {\em ACM computing surveys (CSUR)}, 49(4):1--38, 2017.

\bibitem{shukla2023cisco}
Avinash Shukla, Jalpa Patel, Komal Panzade, and Himanshu Sardana.
\newblock {\em Cisco Cloud Infrastructure}.
\newblock Cisco Press, 2023.

\bibitem{sun2022privacy}
Xin Sun, Chengliang Tian, Changhui Hu, Weizhong Tian, Hanlin Zhang, and Jia Yu.
\newblock Privacy-preserving and verifiable src-based face recognition with cloud/edge server assistance.
\newblock {\em Computers \& Security}, 118:102740, 2022.

\bibitem{sun2016long}
Zhen Sun, Geoff Kuenning, Sonam Mandal, Philip Shilane, Vasily Tarasov, Nong Xiao, et~al.
\newblock A long-term user-centric analysis of deduplication patterns.
\newblock In {\em 2016 32nd Symposium on Mass Storage Systems and Technologies (MSST)}, pages 1--7. IEEE, 2016.

\bibitem{vrable2009cumulus}
Michael Vrable, Stefan Savage, and Geoffrey~M Voelker.
\newblock Cumulus: Filesystem backup to the cloud.
\newblock {\em ACM Transactions on Storage (TOS)}, 5(4):1--28, 2009.

\bibitem{wallace2012characteristics}
Grant Wallace, Fred Douglis, Hangwei Qian, Philip Shilane, Stephen Smaldone, Mark Chamness, and Windsor Hsu.
\newblock Characteristics of backup workloads in production systems.
\newblock In {\em FAST}, volume~12, pages 4--4, 2012.

\bibitem{wang2010secure}
Cong Wang, Ning Cao, Jin Li, Kui Ren, and Wenjing Lou.
\newblock Secure ranked keyword search over encrypted cloud data.
\newblock In {\em 2010 IEEE 30th international conference on distributed computing systems}, pages 253--262. IEEE, 2010.

\bibitem{wang2023svtpm}
Juan Wang, Jie Wang, Chengyang Fan, Fei Yan, Yueqiang Cheng, Yinqian Zhang, Wenhui Zhang, Mengda Yang, and Hongxin Hu.
\newblock Svtpm: Sgx-based virtual trusted platform modules for cloud computing.
\newblock {\em IEEE Transactions on Cloud Computing}, 2023.

\bibitem{wang2019edge}
Shangguang Wang, Yali Zhao, Jinlinag Xu, Jie Yuan, and Ching-Hsien Hsu.
\newblock Edge server placement in mobile edge computing.
\newblock {\em Journal of Parallel and Distributed Computing}, 127:160--168, 2019.

\bibitem{xia2016fastcdc}
Wen Xia, Yukun Zhou, Hong Jiang, Dan Feng, Yu~Hua, Yuchong Hu, Qing Liu, and Yucheng Zhang.
\newblock $\{$FastCDC$\}$: A fast and efficient $\{$Content-Defined$\}$ chunking approach for data deduplication.
\newblock In {\em 2016 USENIX Annual Technical Conference (USENIX ATC 16)}, pages 101--114, 2016.

\bibitem{yang2020data}
Pan Yang, Naixue Xiong, and Jingli Ren.
\newblock Data security and privacy protection for cloud storage: A survey.
\newblock {\em Ieee Access}, 8:131723--131740, 2020.

\bibitem{yang2022secure}
Zuoru Yang, Jingwei Li, and Patrick~PC Lee.
\newblock Secure and lightweight deduplicated storage via shielded $\{$deduplication-before-encryption$\}$.
\newblock In {\em 2022 USENIX Annual Technical Conference (USENIX ATC 22)}, pages 37--52, 2022.

\bibitem{zhu2008avoiding}
Benjamin Zhu, Kai Li, and R~Hugo Patterson.
\newblock Avoiding the disk bottleneck in the data domain deduplication file system.
\newblock In {\em Fast}, volume~8, pages 1--14, 2008.

\end{thebibliography}
\end{document}